\documentclass[]{aa}

\usepackage{graphicx,subcaption,natbib}
\usepackage{txfonts}
\usepackage{lipsum}
\begin{document} 

   \title{Constraints on core-collapse supernova progenitors from explosion site integral field spectroscopy\footnote{Based on observations collected at the European Organisation for Astronomical Research in the Southern Hemisphere under ESO programmes 089.D-0367, 091.D-0482, 093.D-0318, 094.D-0290, and 095.D-0172}}


   \titlerunning{SN progenitor constraints from IFU spectroscopy}
   \authorrunning{H. Kuncarayakti et al.}
   
   \author{
   H. Kuncarayakti\inst{1,2,3,4}\thanks{\email{hanindyo.kuncarayakti@utu.fi}}, 
J. P. Anderson\inst{5},
L. Galbany\inst{6},    
K. Maeda\inst{7,8},
M. Hamuy\inst{4,3},
G.~Aldering\inst{9},  
N.~Arimoto\inst{10,11},    
M.~Doi\inst{12,13},
T.~Morokuma\inst{12},
T. Usuda\inst{14,15}
          }

  \institute{
Finnish Centre for Astronomy with ESO (FINCA), University of Turku, V\"{a}is\"{a}l\"{a}ntie 20, 21500 Piikki\"{o}, Finland
\and
Tuorla Observatory, Department of Physics and Astronomy, University of Turku, V\"{a}is\"{a}l\"{a}ntie 20, 21500 Piikki\"{o}, Finland
\and  
  Millennium Institute of Astrophysics, Casilla 36-D, Santiago, Chile
        \and
             Departamento de Astronom\'ia, Universidad de Chile, Casilla 36-D, Santiago, Chile
\and
European Southern Observatory, Alonso de C\'ordova 3107, Vitacura, Casilla 19001, Santiago, Chile
\and
PITT PACC, Department of Physics and Astronomy, University of Pittsburgh, Pittsburgh, PA 15260, USA
\and
Department of Astronomy, Graduate School of Science, Kyoto University, Sakyo-ku, Kyoto 606-8502, Japan
\and
Kavli Institute for the Physics and Mathematics of the Universe (WPI), The University of Tokyo, 5-1-5 Kashiwanoha, Kashiwa, Chiba 277-8583, Japan
\and
Physics Division, Lawrence Berkeley National Laboratory, 1 Cyclotron Road, Berkeley, CA 94720, USA
\and
Astronomy Program, Department of Physics and Astronomy, Seoul National University, 599 Gwanak-ro, Gwanak-gu, Seoul, 151-742, Korea
\and
Subaru Telescope, National Astronomical Observatory of Japan, National Institutes of Natural Sciences, 650 North A’ohoku Place, Hilo, HI 96720, USA    
\and
Institute of Astronomy, Graduate School of Science, University of Tokyo, 2-21-1, Osawa, Mitaka, Tokyo 181-0015, Japan
\and
Research Center for the Early Universe, Graduate School of Science, The University of Tokyo, 7-3-1 Hongo, Bunkyo-ku, Tokyo 113-0033, Japan
\and
National Astronomical Observatory of Japan, 2-21-1, Osawa, Mitaka, Tokyo, 181-8588, Japan
\and
Department of Astronomical Science, SOKENDAI (The Graduate University for Advanced Studies), 2-21-1 Osawa, Mitaka, Tokyo 181-8588, Japan
             }

   \date{Received; accepted 10 November 2017}

 
  \abstract
   {Observationally, supernovae are divided into subclasses pertaining to their distinct characteristics. This diversity naturally reflects the diversity in the progenitor stars. It is not entirely clear however, how different evolutionary paths  leading massive stars to become a supernova are governed by fundamental parameters such as progenitor initial mass and metallicity.}
   {This paper places constraints on progenitor initial mass and metallicity in distinct core-collapse supernova subclasses, through a study of the parent stellar populations at the explosion sites. 
   }
   {Integral field spectroscopy (IFS) of 83 nearby supernova explosion sites with a median distance of 18 Mpc has been collected and analysed, enabling detection and spectral extraction of the parent stellar population of supernova progenitors.
   From the parent stellar population spectrum, the initial mass and metallicity of the coeval progenitor are derived by means of comparison to simple stellar population models and strong-line methods.
Additionally, near-infrared IFS was employed to characterise the star formation history at the explosion sites.   
   }
   {No significant metallicity differences are observed among distinct supernova types. The typical progenitor mass is found to be highest for supernova type Ic, followed by type Ib, then types IIb and II. Type IIn is the least associated with young stellar populations and thus massive progenitors. 
However, statistically significant differences in progenitor initial mass are observed only when comparing supernovae IIn with other subclasses.
Stripped-envelope SN progenitors with initial mass estimate lower than 25~$M_\odot$ are found; these are thought to be the result of binary progenitors.
   Confirming previous studies, these results support the notion that core-collapse supernova progenitors cannot arise from single-star channel only, and both single and binary channels are at play in the production of core-collapse supernovae.
   Near-infrared IFS suggests that multiple stellar populations with different ages may be present in some of the supernova sites. As a consequence, there could be a non-negligible amount of contamination from old populations, and therefore the individual age estimates are effectively lower limits.
   }
   {}


   \maketitle
%

\section{Introduction}

{A core-collapse (CC) supernova (SN) is produced when the stellar core collapses onto itself at the end of the lifetime of a massive star.}
SNe are one of the brightest phenomena in the Universe. At peak brightness, a SN may rival or even outshine the entire host galaxy where it resides. 
{The explosion distributes heavy elements forged inside the stellar interior \citep{hoyle60,nomoto06}, driving chemical enrichment of the interstellar medium \citep{matteucci86,timmes95}, and may trigger new waves of star formation \citep{mccray87,thornton98}.}
Thus, SNe are important players in cosmic evolution.

A star needs to be massive enough to experience core collapse at the end of its life. It has been established that in general stars with initial mass around 8 $M_\odot$ and above finish their lifetime with a CC (see reviews by \citealt{langer12}, and \citealt{smartt09araa}, from both theoretical and observational standpoints).
One lingering question is how does the observed diversity in SNe relate to the distribution of massive star parameters in terms of initial mass and pre-SN evolutionary state, or in other words: \emph{which kind of massive star gives rise to which type of CCSN?}

Some of the massive star progenitors of SNe have been directly identified in archival images taken before the explosion. This method provides currently the most reliable constraints on the physical parameters of SN progenitors.
The hydrogen-rich type-II SNe, which are the most abundant among CCSN subtypes \citep{li11}, are found to be predominantly produced by red supergiant (RSG) progenitors \citep{smartt09}. These SN II progenitors are born with initial masses between $\sim$8-20 $M_\odot$ and spend their main sequence lifetime as late-O/early-B spectral type stars. By the end of stellar evolution, they still retain most of their H envelope and appear as RSGs, before eventually exploding as type-II SNe.

The H-poor CCSNe\footnote{Also known as the stripped-envelope (SE) SNe.}, which encompass types Ib (H spectral lines absent, He present), Ic (both H and He absent), and IIb (He present, little H) \citep{filippenko97}, are thought to arise from massive stars that have lost most of their outer envelope due to some mechanism. 
One way for a massive star to lose its outer envelope is through metallicity-driven stellar winds. With higher metallicity, the stellar wind becomes stronger due to increased line opacity, thus the rate of mass removed from the star would be higher \citep[$\dot{M} \propto Z^{\textrm{0.6-0.7}}  $][]{vink01}. 
The progenitors of SESNe are thought to {be} classical Wolf-Rayet (WR) stars that are deprived of the envelope and exhibit strong stellar winds. At solar metallicity, the lower limit of the initial ZAMS (zero-age main sequence) mass for a star to eventually evolve into a WR star is around 25 $M_\odot$ \citep{crowther07}.

An alternative to the metallicity-driven wind mechanism is mass loss via close binary interaction \citep{podsiadlowski92}. In this scenario the SN progenitor star is in a close binary system, and as stellar evolution proceeds, the progenitor expands and overfills its Roche lobe---resulting in mass transfer and significant mass loss. With such a mechanism, the star does not need to be initially as massive as WR star progenitors ($\gtrsim$25 $M_\odot$) in order to lose the outer envelope and eventually explode as a SESN. 
{The detections of SESN progenitors in pre-explosion images in the recent years seem to support this binary scenario.}

Five SN IIb progenitor candidates have been identified in pre-explosion images: SNe 1993J, 2008ax, 2011dh, 2013df, 2016gkg. It is very interesting that for all these cases, 
{the observations seem to be consistent with the binary progenitor scenario (\citealt{maund04,folatelli14,folatelli15,maeda15,kilpatrick17}, although in some cases single progenitor scenario cannot be ruled out, \citealt{maund15,sravan17}).}
For these cases, the derived progenitor mass is typically higher than most SN II progenitors ($\sim$15 $M_\odot$).

In contrast to SN II and IIb progenitors, efforts to detect the progenitors of SNe Ib and Ic have proved unfruitful \citep[see e.g.][]{eldridge13}. Currently there is only one case of SN~Ib/Ic progenitor detection: that of iPTF13bvn, a SN Ib \citep{cao13}. The progenitor star was constrained to be a sub WR-mass star ($M_\textrm{ZAMS} < 25$ $M_\odot$) in a massive binary system through a number of independent methods \citep{bersten14,hk15,folatelli16,eldridge16,hirai17}. Along with the observational constraints on SN IIb progenitors this questions the importance of massive, single WR stars in the production of SESNe. 

An independent way to constrain SN progenitors is by studying their environments. Unlike direct detection methods that are strictly limited to the availability of useful archival images, and analyses based on {SN light curves and spectra} that are restricted to the time window when the SN is still visible, environment studies do not depend on these limitations and still can be achieved much later after the SN has faded. Therefore, it is possible to build statistically large samples of SN environments and derive strong progenitor constraints from there \citep[see][for a review]{anderson15}. From the environments, the estimate for metallicity and age of the stellar population that gave rise to the SN can be derived. This subsequently reflects the birth metallicity and mass of the SN progenitor, which are the two fundamental factors driving massive star evolution up to the terminal SN explosion \citep[see e.g.][]{heger03,georgy09}.

The birth mass and metallicity of SN progenitors are thought to strongly affect the mass loss, and thus the degree of envelope stripping and the eventual SN type.
Several studies have used proxies such as host central metallicity and SN radial distance from host centers for metallicity estimates of SN progenitors \citep[e.g.][]{prieto08,anderson09}. A more direct approach to measure metallicity is by obtaining the spectrum of the explosion site, and using strong line diagnostics to derive metallicity. 
\citet{anderson10} and \citet{leloudas11} showed that the gas-phase metallicities measured at SN explosion sites do not show statistically significant differences for SNe of different types, while exhibiting an overall trend of higher metallicity for SESN explosion sites. \citet{modjaz08} showed that the environments of broad-lined type-Ic SNe (IcBL) that are associated with long gamma-ray bursts (GRB) are more metal-poor compared to their GRB-less counterparts. 

If SESNe are produced by classical WR stars that are single and massive ($\gtrsim$25 $M_\odot$) at birth, there may be a correlation between birth mass and degree of envelope stripping, such that the most massive stars would lose H and He more easily compared to their lower mass counterparts. This scenario implies that SESNe are, on average, more massive than SNe II. \citet{anderson12} investigated this issue using a pixel statistics technique \citep{fruchter06,james06} that essentially constrains the association of SNe with H$\alpha$ emission from a nearby H \textsc{ii} region. As stars age, they drift further from the star-forming region birthplace and show less association to the H$\alpha$ emission. Thus, more massive stars (including SN progenitors) should show higher association to star formation compared to the lower mass ones. It was found that indeed SNe Ic show the highest association with star formation, thus highest progenitor mass, followed by SNe Ib that somewhat resembles the SNe II distribution. 
\citet{kangas17} went further with this technique and compared the pixel statistics of SNe with that of main sequence and evolved stars. They confirmed the finding of \citet{anderson12}, and suggested the following pairs of SN types and progenitors: SNe Ic are consistent with WN stars $\gtrsim 20$ $M_\odot$, while SNe Ib and II occupy the lower mass range between around 9 and 15 $M_\odot$.


With the recent deployment of integral field spectroscopy (IFS/IFU spectroscopy), the field of SN environments has received a significant boost. IFS allows the collection of both spatial and spectral information of the environments in an efficient manner, while at the same time returning a large amount of information. With IFS, the spectra of the exact explosion site, the immediate local surroundings, and (depending on the size of the IFU field of view/FoV) the host galaxy, can be obtained simultaneously. Thus, in contrast to slit spectroscopy, IFS offers the possibility of identifying and extracting the parent stellar population of the SN rather than simply integrating over the slit aperture \citep{hk13a,hk13b}, and also comparing its properties to the wider environment and the rest of the host galaxy \citep{galbany14,galbany16,galbany16muse,galbany17,kruehler17}. 

In this paper we aim to refine the work presented in \citet{hk13a,hk13b}, in constraining the mass and metallicity of SN progenitors by way of spectral analysis of the parent stellar population. Similarly, IFU spectroscopy of the SN explosion site was utilised to identify the parent population and extract the spectrum, from which the SN progenitor properties were derived assuming coevality. In total, 24 SN sites were used in \citet{hk13a,hk13b}, and with the new dataset presented in this study, this number has now increased more than threefold {(83 SN sites)}. 
Furthermore, the current study involves other CCSN subtypes not covered in \citet{hk13a,hk13b}: SNe IIb, IIn, and IcBL, and thus offers a more complete understanding towards massive star evolution and the endpoint SNe.

Following the introduction, the observations and data reductions are presented in Section \ref{sec:obs}. Analyses on progenitor metallicity, initial mass, and star formation history at the explosion site are presented in Section \ref{sec:ana}, accompanied with discussions in Section \ref{sec:discu}. Section \ref{sec:conc} concludes the paper.

\section{Observations and data reductions}
\label{sec:obs}
\subsection{Sample description}

The SNe used in this work were selected from the Asiago Supernova Catalog \citep{barbon99}. 
The following selection criteria were used:
\begin{itemize}
\item Non-thermonuclear (type-Ia) SNe.
\item $\delta \le 30^\circ$, to enable observations from the southern hemisphere.
\item Host galaxy redshift $cz \le 3000$ km s$^{-1}$, to limit the survey volume.
\item Host galaxy inclination $\le 65^\circ$ (following \citealt{crowther13}), to minimize chance superposition and other effects that may be attributed to a line of sight through the host galaxy disk. 
\item Only relatively recent SNe, discovered not earlier than 1970, and not after January 2013 were selected. This is to ensure reliable typing/position and minimal contamination from the late-time SN emissions.\footnote{ 
One SN (2017ahn, type II) was discovered much later after the IFU observations \citep{tartaglia17}. The explosion site of this SN was serendipitously covered in the IFU data, and therefore it was included in the sample.}
\end{itemize}

New IFU data of 62 SNe following these criteria were obtained between 2014-2015. One object, SN 1992bd (type II), was not considered in the analysis due to its position being very close to the host galaxy center, which contains an active nucleus that makes contamination-free extraction of the parent stellar population light difficult. 
Twenty-two\footnote{SNe 1961I and 1964L were omitted as they do not satisfy the $\geq 1970$ criterion.} SNe from \citet{hk13a,hk13b} were added to this sample to construct the total sample of 83 SNe (see Table \ref{tab:obj}).
In the subsequent analyses, all non-interacting H-rich type-II SNe are collectively put under one group of SN II, without considering the traditional subtypes of IIP and IIL as these hydrogen-rich explosions seem to comprise a continuum without obvious dichotomy in the light curve shape \citep{anderson14,sanders15,galbany16cats}.

\begin{figure}
   \centering
   \includegraphics[width=\hsize]{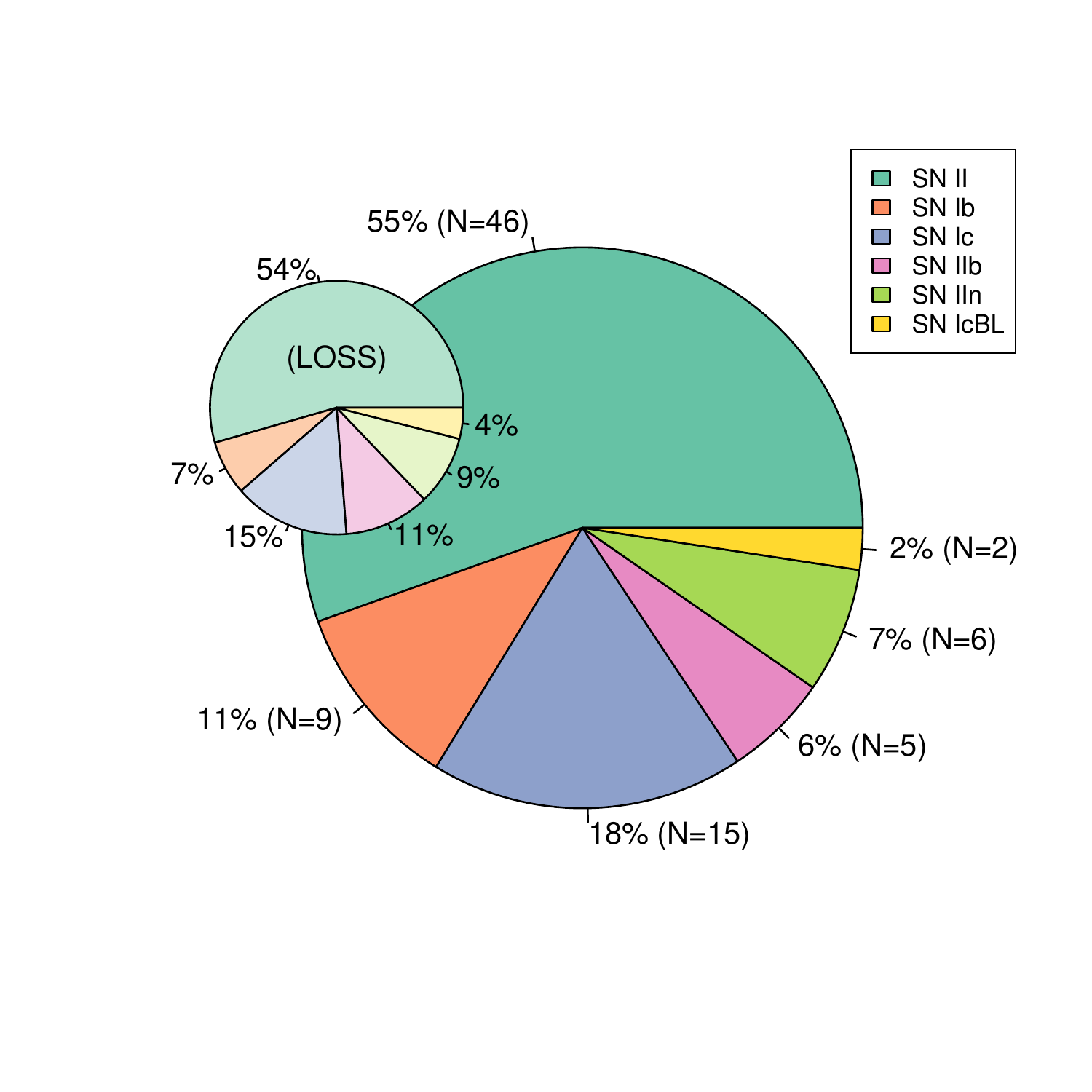}
      \caption{Relative frequency of each SN type in the current sample (large pie chart), compared to that of LOSS \citep[][small pie chart with pale colours]{smith11}.}
         \label{pie}
   \end{figure}

Figure~\ref{pie} shows the pie chart of relative frequency of each SN subtype in the sample. For comparison, the relative frequencies of SNe from the LOSS survey \citep{smith11} are also shown. The relative frequency of SNe in both samples are consistent when the sample is divided into hydrogen-poor and hydrogen-rich. Within the hydrogen-poor SN subset itself there are differences, but these are not statistically significant.

\begin{figure}
   \centering
   \includegraphics[width=\hsize]{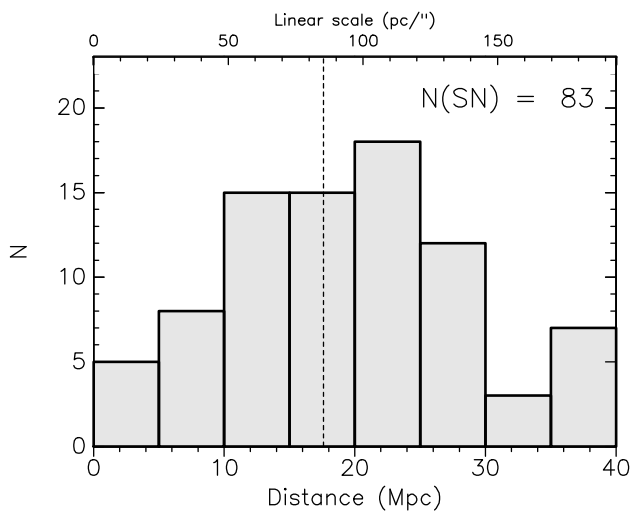}
      \caption{Histogram of SN host distance. The upper abscissa indicates the projected linear scale in pc/". The vertical dashed line indicates the median value.}
         \label{histo_d}
   \end{figure}

Figure \ref{histo_d} shows the histogram of the SN host distances. All of the objects are closer than 40 Mpc, with 90\% within $\sim$30 Mpc and a median distance of 17.6 Mpc. At these distances, the typical spatial resolution in the optical data as constrained by natural seeing is better than 100 pc.
To date, the current study has the shortest median distance and thus the highest spatial resolution compared to previously published studies focusing on SN environments \citep[see e.g.][]{anderson10,anderson15,
leloudas11,modjaz11,sanders12,
galbany14,galbany16}.

\subsection{Optical observations and data reductions}

All optical observations for the new dataset were conducted with the Very Large Telescope (VLT) at the Cerro Paranal Observatory, Chile, in classical visitor mode. The instruments VIMOS \citep{lefevre03} and MUSE \citep{bacon14} were used for this purpose.

VIMOS was used in IFU mode and with a medium resolution (MR) grating, while MUSE was used in the Wide Field Mode. 
VIMOS gives 0.33"/spaxel scale within a 13"$\times$13" FoV, while MUSE gives 0.2"/spaxel within a 1'$\times$1' FoV.
Table \ref{tab:ifu} lists the instrument configurations and the resulting spatial and spectral characteristics.
The observations spanned around one year within 2014-2015, in five different runs in 2014 Apr, 2014 Jul, 2014 Nov, 2015 Feb (VIMOS\footnote{ESO Programme ID 093.D-0318 and 094.D-0290.}), and 2015 May (MUSE\footnote{ESO Programme ID 095.D-0172.}). 

Table \ref{tab:obj} lists the IFU instruments used for each object, and the typical seeing size during the exposure. The sky conditions during the observations were generally clear, with some conducted in lightly cloudy conditions. Spectrophotometric standard stars were observed for the purpose of flux calibration.

VIMOS observations were done in $2 \times 1800$ s on-source exposures, targeting the SN sites. In MUSE observations, the whole SN host galaxy was observed, taking advantage of the wide field of view. In several instances multiple pointings were taken due to the large extent of the galaxy. For each MUSE pointing, four dithered positions of 450 s exposure time were integrated on-source resulting in a total exposure time of 1800 s per pointing. A separate sky pointing was also observed for each object for the purpose of background sky subtraction.

Raw data reduction was carried out using VIMOS and MUSE data reduction pipelines, run using the Reflex interface \citep{freudling13}. For both VIMOS and MUSE raw data, the reduction procedures of bias subtraction, {flat-fielding}, wavelength and flux calibration were applied, and finally cube reconstruction was performed. MUSE cubes were sky subtracted using the blank sky pointings and further corrected for atmospheric effects using the Zurich Atmospheric Package \citep[ZAP][]{soto16}. Sky subtraction in VIMOS data was done in the subsequent spectral extraction step, where the sky background was defined with an annulus around the object extraction aperture.

From the IFU datacube, the parent stellar population of the SN was identified.
Physically these are young stellar clusters/H~\textsc{ii} regions lying within one seeing radius from the SN positions.
{The SN coordinates were obtained from the Asiago SN database, and these were used to localise the SN position in the datacube. The World Coordinate System (WCS) of the datacube was generated by the reduction pipeline, and reflects the telescope pointing accuracy.}
{The accuracy of the WCS coordinates was around 1"-1.5". This was determined using foreground stars and several SNe IIn that were still visible.
Whenever possible, the SN position in the datacube was further checked with available broad-band images of the SN or reported offsets towards the centre of the host galaxy.
}
A one-dimensional spectrum of the stellar population was extracted from the datacube by using an aperture, whose radius is set to the size of seeing FWHM.
Visualization and extraction were done using the QFitsView\footnote{\href{http://www.mpe.mpg.de/~ott/dpuser/qfitsview.html}{http://www.mpe.mpg.de/~ott/dpuser/qfitsview.html}} \citep{ott12} software. Analysis of the 1-dimensional spectrum was carried out using IRAF\footnote{\textsc{Iraf} is distributed by the National Optical Astronomy Observatory, which is operated by the Association of Universities for Research in Astronomy (AURA) under cooperative agreement with the National Science Foundation.}, as described in the next section.

\subsection{Infrared observations and data reductions}

The optical IFU data are supplemented with near-infrared IFU observations, primarily for the purpose of constraining star formation history (see \ref{sec:sfh}). 

VLT/SINFONI \citep{eisenhauer03,bonnet04} was used in the K-band with the 0.1"/spaxel scale, giving a field of view of $3" \times 3"$ (Table~\ref{tab:ifu}). Adaptive optics were used with either natural guide stars, or laser guide stars when applicable, resulting in near diffraction-limited observations.
The observations were conducted between 2012--2015, in several runs in both visitor and service modes\footnote{ESO Programme ID 089.D-0367, 091.D-0482, 093.D-0318, 094.D-0290.}. 
The typical on-source total integration time per object is 6900 s, taken in 300 s dithers for both object and sky frames.
Ten SN sites from the optical sample were observed.

The raw data were reduced using the SINFONI instrument pipeline within the GASGANO interface, which includes the standard procedures of bias subtraction, flatfielding, distortion correction, wavelength and flux calibration, and sky subtraction. 
The resulting datacubes were subsequently analysed using QFitsView and IRAF.
{The uncertainty of the SN position in the SINFONI datacube was estimated to be similar to the 1"-1.5" uncertainty in VIMOS/MUSE datacubes, as the instruments use the same VLT telescope. Due to the narrow SINFONI FoV, no foreground stars could be used for checking the WCS accuracy. In some cases, the general appearance of the SN field in the K-band image generated from the datacube were compared to the optical image (see Section \ref{sec:sfh}), and with this the aforementioned estimate was confirmed.}

\begin{table*}
\caption{Instrument configurations used in this study. \label{tab:ifu}}
\label{table:1}      
\centering          
\begin{tabular}{lcccc}
\hline \hline
Instrument    & Spaxel size & IFU FoV & Wavelength coverage & Spectral dispersion \\
\hline
VLT/VIMOS & 0.33" & 13" $\times$ 13" & 4850--10000 \AA & 2.6 \AA/pixel \\
VLT/MUSE & 0.2" & 60" $\times$ 60" & 4650--9300 \AA & 1.25 \AA/pixel \\
UH2.2m/SNIFS & 0.43" & 6.4" $\times$ 6.4" & 3300--9300 \AA & 2.2 (blue arm), 3.0 (red arm) \AA/pixel \\
Gemini-N/GMOS & 0.2" & 5" $\times$ 7.5" & 4000--6800 \AA & 0.45 \AA/pixel \\
VLT/SINFONI & 0.1" & 3" $\times$ 3" & 1.95--2.45 $\mu$ & 2.45 \AA/pixel \\
\hline
\end{tabular}
\end{table*}

\section{Analysis and results}
\label{sec:ana}
\subsection{Metallicity}
\label{sec:metal}

Gas-phase abundance of oxygen is used as the proxy for metallicity. The widely used N2 index was used to calculate 12+log(O/H), according to the calibration by \citet{marino13}. 
Due to the unavailability of the H$\beta$ line in the data, the O3N2 index was only available for a part of the sample thus the N2 index was used for the whole sample.
N2 uses the ratio of emission lines H$\alpha$ and [N \textsc{ii}] $\lambda6584$, which are closely spaced in wavelength and thus robust against uncertainties introduced by reddening or flux calibration.
As the 12+log(O/H) values from \citet{hk13a,hk13b} are originally in the \citet{pp04} scale, the corresponding 12+log(O/H) values on \citet{marino13} scale were calculated and used.
The solar oxygen abundance was taken as 12+log(O/H)$_\odot$ = 8.69 \citep{asplund09}.

The H$\alpha$ and [N \textsc{ii}] $\lambda6584$ lines were measured in the one dimensional spectrum by fitting a Gaussian curve, after subtracting the stellar continuum using a polynomial function. 
This approach does not make assumptions on the underlying stellar population behind the gas, therefore it does not take into account the underlying stellar absorption.
The resulting H$\alpha$ and [N \textsc{ii}] line strengths are then directly used for the calculations of the N2 index and subsequently metallicity. In several cases, no emission was detected at the explosion sites, or the SN was still bright at the time of the observation\footnote{This late-time emission occurs for a number of interacting type-IIn SNe, where CSM interaction is still in progress  and resulting in an enduring light display. SN 1978K was found to be still bright at 36 years after the explosion, as was discussed in a separate paper \citep{hk16a}.}. For such cases, the nearest H~\textsc{ii} region spectrum is extracted and the metallicity estimate was adopted for the metallicity value at the explosion site. This procedure is only applicable for SN sites observed with MUSE, due to the large field of view allowing the observation of the whole host galaxy (SNe 1978G, 1983K, 1988E, 2011fh). 
In these cases, the average projected offset from the SN positions to the nearest H \textsc{ii} region is 540 parsec.

In addition, metallicity from the calibration of \citet{dopita16} was also calculated. Their new calibration similarly uses H$\alpha$ and [N \textsc{ii}] $\lambda6584$\footnote{The [N~\textsc{ii}] line wavelength was incorrectly written as 6484 \AA~ in the paper (Dopita et al., priv. comm.).} lines, and additionally [S \textsc{ii}] $\lambda\lambda6717,6731$.
The results of metallicity analysis are unchanged when this calibration is applied.

\begin{figure*}
\centering
\begin{subfigure}{.5\textwidth}
  \centering
  \includegraphics[width=\linewidth]{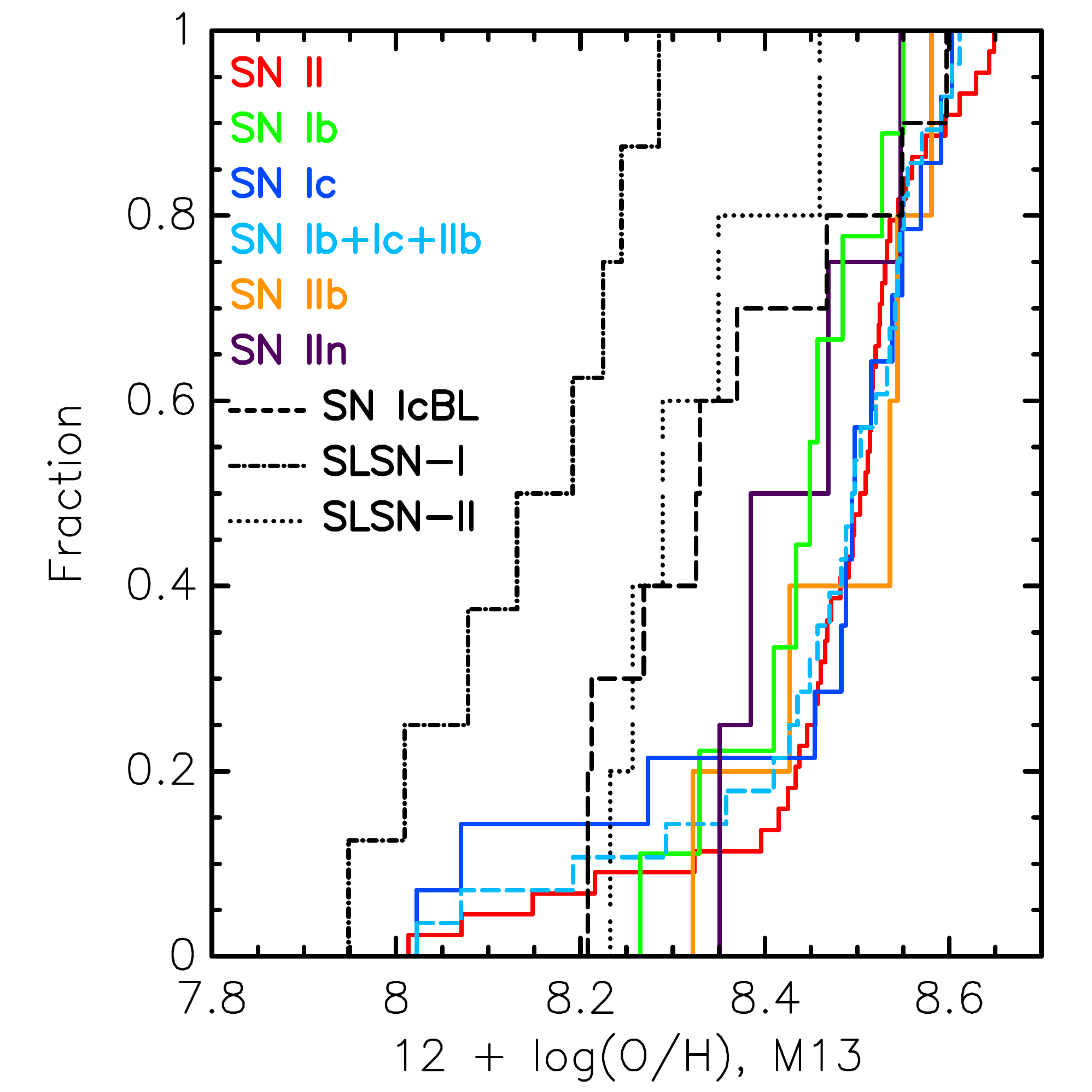}
\end{subfigure}%
\begin{subfigure}{.5\textwidth}
  \centering
  \includegraphics[width=\linewidth]{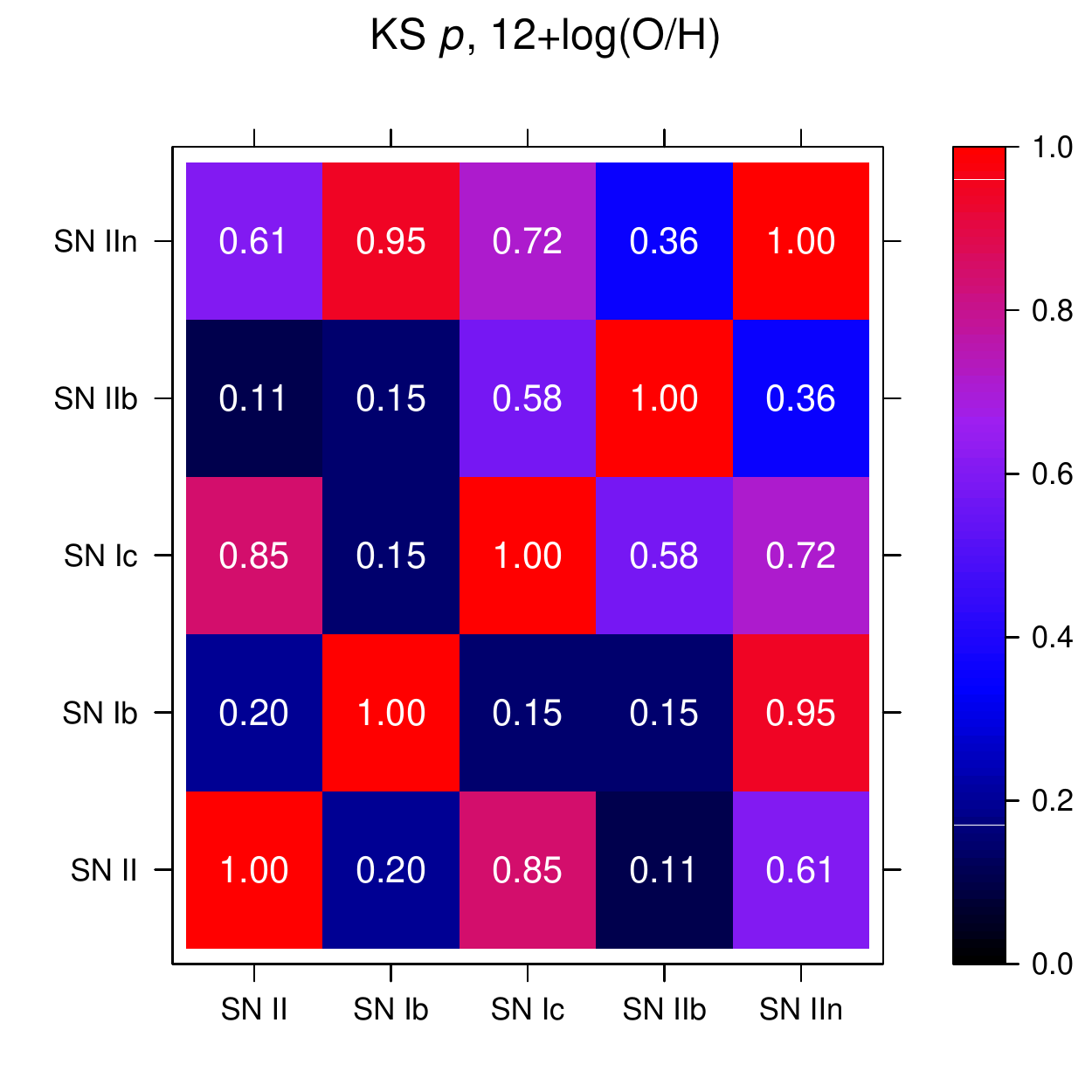}
\end{subfigure}
\caption{
\textit{(Left)} Observed cumulative distributions of 12+log(O/H) for different SN subtypes. SNe Ic are colour-coded blue, Ib green, IIb orange, IIn purple, II red, and SESNe (Ic+Ib+IIb combined) dashed cyan. For comparison, the literature distributions of SN IcBL (dashed line), SLSN-I (dash-dotted line), and SLSN-II (dotted line) are also shown.
\textit{(Right)} Matrix for the K-S test result between different SN subtypes. The colour bar indicates the probability that samples were drawn from the same metallicity distribution, where the black end indicates significantly different distributions.
\label{edf_z}}
\end{figure*}

The left panel of Figure~\ref{edf_z} shows the cumulative distribution function of metallicity of different SN types. The SNe are colour-coded according to their subtypes. 
The distributions do not show any particular pattern and are practically superposed one on top of another. The lowest-metallicity bin of the sample is occupied by a number of SNe II and Ic only, between 12+log(O/H) = 8.0 and 8.2 dex.
Metallicity measurements for extreme SNe such as superluminous (SL) SNe \citep{leloudas15} and SNe Ic-BL \citep{modjaz11}, both converted to the \citet{marino13} scale, are also shown for comparison. It is readily evident that the metallicity of normal CCSN events is typically higher compared to those. The two SNe Ic-BL in our sample, SNe 1998bw and 2009bb, have 12+log(O/H) metallicities of 8.30 and 8.49 dex, respectively.

Two-sample Kolmogorov-Smirnov (K-S) statistical tests were performed to analyse whether two different SN types arise from the same population. The right panel of Figure \ref{edf_z} shows the matrix for the K-S test result. None of the compared SN subtype pairs show significant difference, with K-S $p < 10\%$. 

\subsection{Progenitor age and initial mass}
\label{sec:age}

Stellar evolution is constrained by the amount of nuclear fuel available for burning in the stellar core. More massive stars burn fuel more rapidly compared to their less massive counterparts. 
Within a coeval stellar population, stars that were born with the highest mass have the shortest lifespans and die out first, leaving behind stars with lower masses. 
Such a simple stellar population is formed out of a single, homogeneous molecular cloud in an instantaneous burst of star formation, hence all the stars share the same age and metallicity. Therefore, by determining the age of the stellar population it is possible to estimate the age, thus lifetime, of the last star that died out. 
This star was observed as the SN, and all the current remaining stars in the stellar population must have initial mass lower than that of the SN progenitor star.

In this work, as in \citet{hk13a,hk13b}, the stellar population that is present at the SN explosion site is assumed to be the parent stellar population from which the SN progenitor emerged. The age of the stellar population is taken as the SN progenitor lifetime span, and this is subsequently converted into an initial mass estimate. 
Due to a number of possible factors this initial mass estimate is effectively an upper limit, i.e., the actual SN progenitor masses are more likely to be lower than those derived from stellar population age. For example, including continuous star formation or binary population may extend the lifetime of H$\alpha$ emission used in age determination, thus the stellar population age derived with such assumptions will be considerably older compared to the instantaneous or single-star ones for a given H$\alpha$ equivalent width (EW)
\citep[see e.g.][]{leitherer99,eldridge09,crowther13}.

In this study, the age indicator H$\alpha$EW is used to constrain the stellar population age. The indicator is measured as the ratio between the emission line and continuum fluxes at H$\alpha$.  
It is effectively a measure of the number ratio between the ionizing stars of OB spectral class responsible for the H$\alpha$ emission line and the other non-ionizing (lower mass) stars in the stellar population, assuming a fixed IMF. As the stellar population ages, these ionizing stars will be reduced in number while the number of lower-mass stars remains constant, therefore resulting in the decline of H$\alpha$EW. 

Measurements of H$\alpha$EW were performed on the 1-D spectra of the stellar population, using IRAF/\textit{splot}. The spectral continuum was normalized using a polynomial function and the emission line was fit using a Gaussian function. 
As in the metallicity estimate, this measurement involves a very small window in wavelength space, thus is robust against uncertainties introduced by reddening and flux calibration. 

\begin{figure}
   \centering
   \includegraphics[width=\hsize]{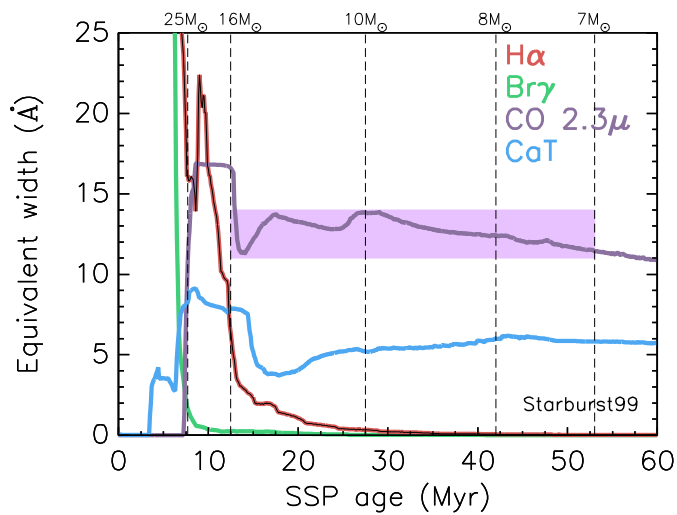}
      \caption{The evolution of Starburst99 age indicators, shown for $Z = 0.02$. The upper abscissa represents the stellar lifetime at the corresponding SSP age. H$\alpha$ is indicated with red, Br$\gamma$ green, CO 2.3 $\mu$ purple, and CaT blue. 
      The CO EW values where typical SN II progenitors ($M_\textrm{ZAMS} \approx$ 8-16 $M_\odot$) are expected to be found are encompassed within the purple shaded area.}
         \label{ageindic}
   \end{figure}

Simple stellar population (SSP) models from Starburst99 \citep{leitherer99} were used to compare the observed age indicators and infer the age of the stellar population. 
SSP models have been shown to be reliable for analysing young stellar populations \citep{hk16b}.
While H$\alpha$EW is the primary age indicator used in this study, a number of other spectral lines are also useful as age indicators. 
{These include the EWs of Br$\gamma$ and CO $\nu=$ 2-0 band (the 2.3 $\mu$m overtone; hereafter CO 2.3 $\mu$) in the K-band (see section~\ref{sec:sfh}), }
and also to a limited extent the near-infrared Ca-triplet lines \citep[see][]{hk13a,hk13b}. The observed EW is compared to the SSP model at the corresponding metallicity to estimate the stellar population age. Figure~\ref{ageindic} shows the evolution of these age indicators. 

\begin{figure*}
\centering
\begin{subfigure}{.5\textwidth}
  \centering
  \includegraphics[width=\linewidth]{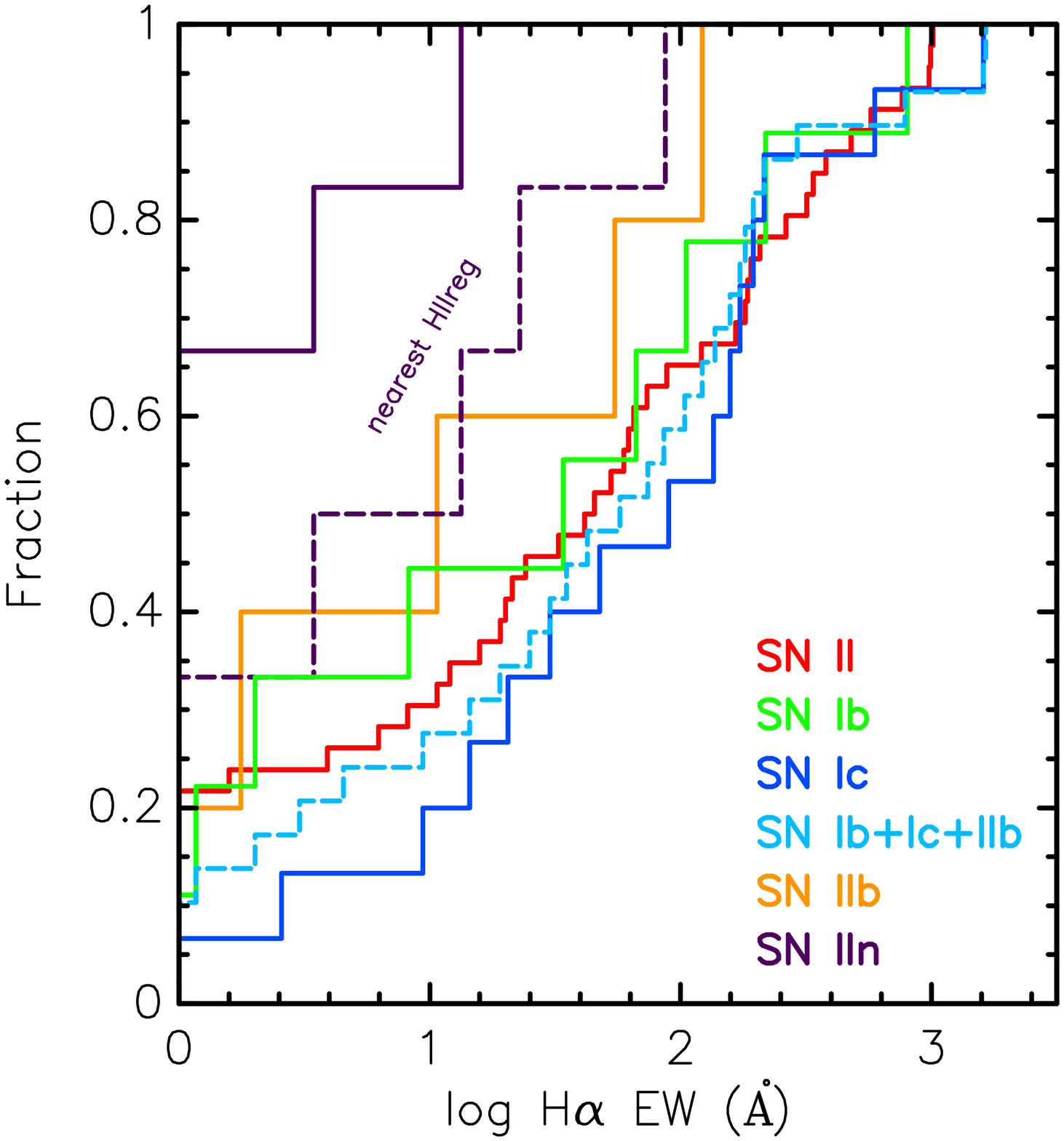}
\end{subfigure}%
\begin{subfigure}{.5\textwidth}
  \centering
  \includegraphics[width=\linewidth]{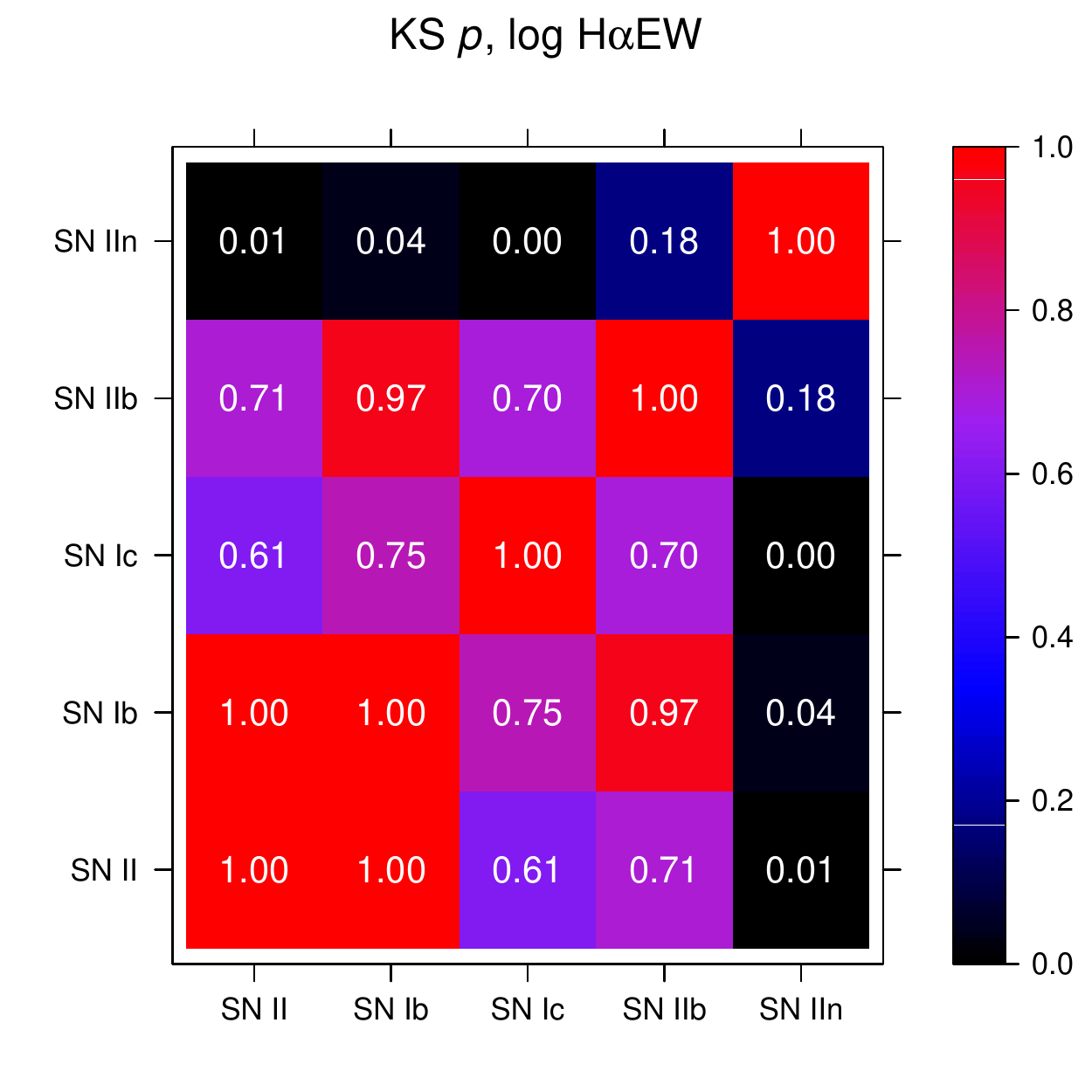}
\end{subfigure}
\caption{\textit{(Left)} Observed cumulative distribution of log(H$\alpha$EW) for different SN subtypes. SN sites which show no H$\alpha$ emission are plotted as having log(H$\alpha$EW) = 0. The dashed purple line indicates SN IIn distribution using the nearest H \textsc{ii} regions (see text for description).
\textit{(Right)} Matrix for K-S test result. 
Colour codes in the figures are the same as in Figure~\ref{edf_z}.
\label{edf_haew}}
\end{figure*}

The observed H$\alpha$EW corresponds to the age of the stellar population, through SSP models. Starburst99 SSP models in the corresponding metallicities ($Z = 0.004, 0.008, 0.02$) were used, assuming single stars born in an instantaneous star formation and distributed in mass according to Salpeter IMF.
The cumulative distribution of the measured H$\alpha$EW of the SN parent populations is given in the left panel of Figure~\ref{edf_haew}. This is the empirical measure of the SN progenitor lifetime in its rawest form. In the figure it is apparent that the SN IIn distribution prefers low H$\alpha$EW, followed by SN IIb, then SN Ib and SN II which are quite closely separated, and finally the SN Ic distribution that shows a preference for high H$\alpha$EW. The majority of SN IIn explosion sites do not show any H$\alpha$ emission, while nearly all SN Ic sites have significant H$\alpha$ emission. Even if the next nearest H \textsc{ii} regions to the SNe IIn are used for the H$\alpha$EW measurement (see section \ref{sec:metal}), it is still apparent that SNe IIn have the lowest H$\alpha$EW among the SN types. The result of the K-S test in the right panel of Figure~\ref{edf_haew} confirms that SNe IIn are statistically different from most of the other SN types. On the other hand, the differences between the other SN types in terms of parent population H$\alpha$EW are not statistically significant.

\begin{figure*}
\centering
\begin{subfigure}{.32\textwidth}
  \centering
  \includegraphics[width=\linewidth]{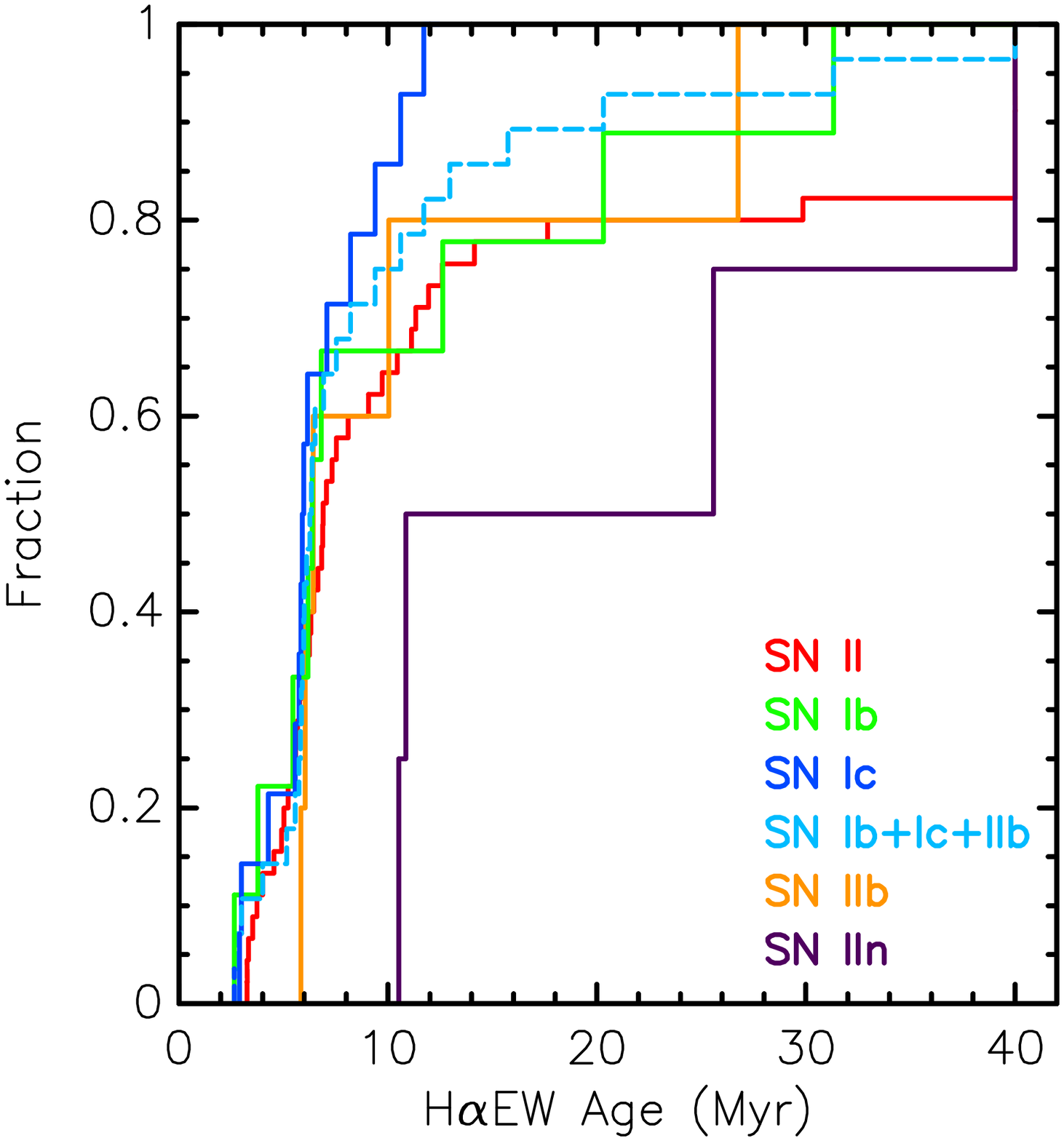}
\end{subfigure}%
\begin{subfigure}{.32\textwidth}
  \centering
  \includegraphics[width=\linewidth]{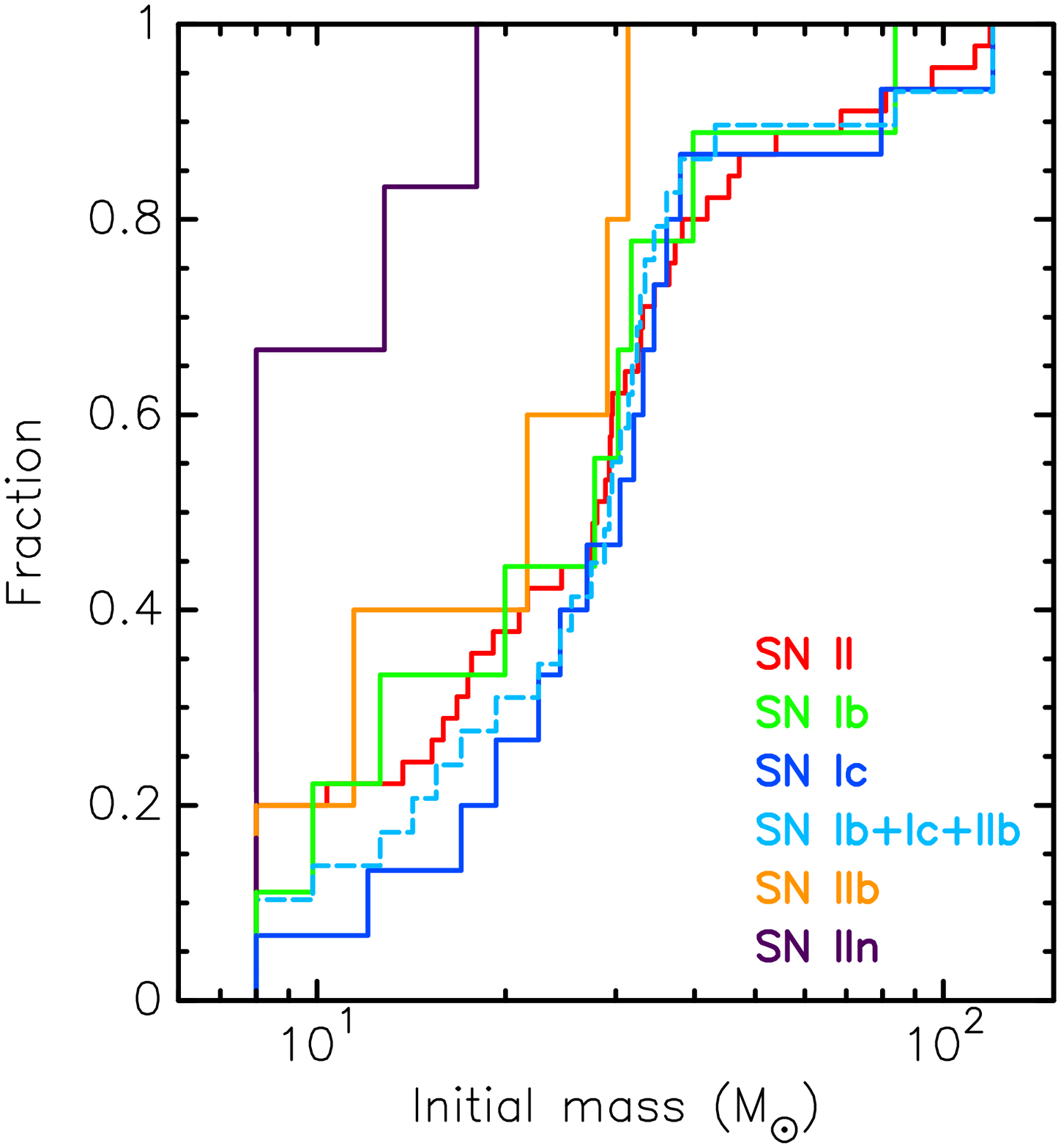}
\end{subfigure}
\begin{subfigure}{.3\textwidth}
  \centering
  \includegraphics[width=\linewidth]{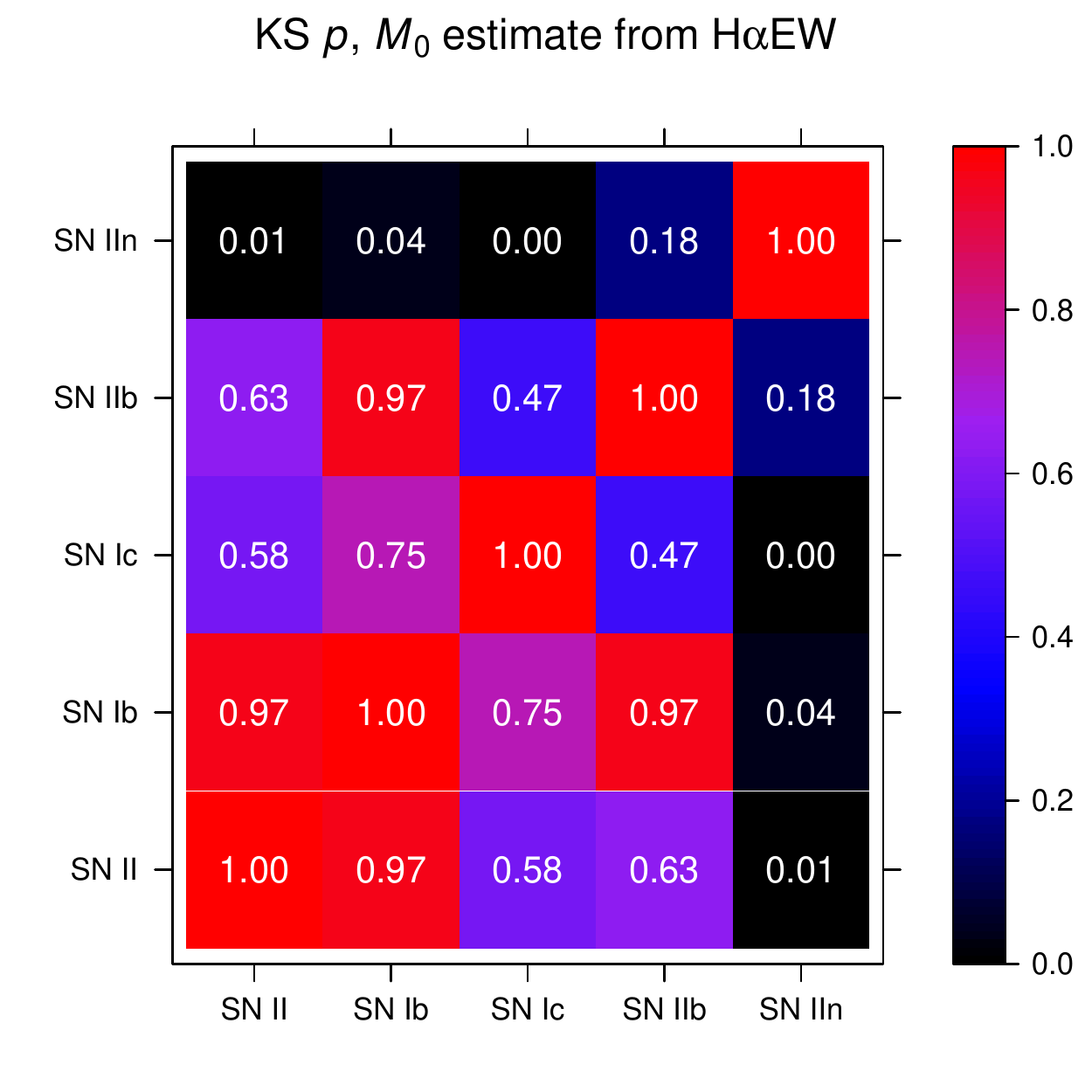}
\end{subfigure}
\caption{\textit{(Left and center)} Observed cumulative distribution of parent stellar population age and progenitor initial mass for different SN subtypes. 
SN sites which show no H$\alpha$ emission are plotted as having the age of 40 Myr, corresponding to the single-star SN progenitor lower mass limit of 8 $M_\odot$.
\textit{(Right)} Matrix for K-S test result.
Colour codes in the figures are the same as in Figure~\ref{edf_z}.
\label{edf_m}}
\end{figure*}

The parent population age, in turn, translates into the age of the previous most massive star in the population which exploded as a SN. This essentially is the lifetime of the progenitor star, which is governed by its initial mass. Stellar evolution models by \citet[][for $Z = 0.02$]{bressan93} and \citet[][for $Z = 0.008, 0.004$]{fagotto94} were used to derive the star initial mass from the lifetime. Figure \ref{edf_m} shows the cumulative distribution of the derived age and initial mass for different SN subtypes. The similar trend as in Figure \ref{edf_haew} is observed: SNe Ic occupy the young age and high mass end, while on the other side SNe IIn are characterised by old age/low mass, and in between the SNe II, IIb, and Ib together show rather similar distributions in age and initial mass. Again, the differences between the SN subtypes are not found to be statistically significant, except in the case of SNe~IIn.

\subsection{Star formation history}
\label{sec:sfh}
As discussed in \citet{hk13a,hk13b}, one of the most important caveats in the determination of stellar population age is the uncertainty in the star formation history. The age-dating of the stellar population is based on the assumption that star formation was instantaneous, while it is uncertain whether this is actually the case. What is observed is the combined light of many different stars and interstellar gas clouds, and it is not possible to ascertain the actual history of star formation in the region. 
However, using a combination of age indicators it is possible probe distinct stellar populations of different ages, thus giving clues whether the star formation was instantaneous or not.

Using near-infrared (NIR) IFU data obtained with SINFONI, here we attempt to alleviate this problem. The age indicators Br$\gamma$ emission EW and CO 2.3 $\mu$ absorption EW probe entirely different SSP ages. Br$\gamma$ exclusively probes stellar populations younger than $\sim$ 8 Myr, while CO probes the population older than that (Figure~\ref{ageindic}). Moreover, CO EW values between 11-14 $\AA$ indicate stellar population ages corresponding to the typical SN II progenitor mass of 8-16 $M_\odot$ obtained from direct progenitor detection \citep{smartt09}.
If both indicators are observed to be present at the same location, it may suggest that the star formation in the area did not proceed in one single burst (i.e., there are two distinct populations with average ages of $<8$  and $>8$ Myr). This immediately separates the regions with single or multiple bursts of star formation. Although the actual star formation history is not recovered, being able to assess whether the star formation was single burst or not is an important point for the current work. 
{In addition, this dividing line at $\sim$8 Myr corresponds to the lifetime of $\sim$25 M$_\odot$ stars, which is thought to be the lower limit of the ZAMS mass of single WR stars \citep{crowther07}.}
Thus, the detection of strong Br$\gamma$ emission with EW corresponding to $\lesssim 8$ Myr may be used to constrain the presence of such massive stars.

\begin{figure*}
\centering
  \includegraphics[width=\linewidth]{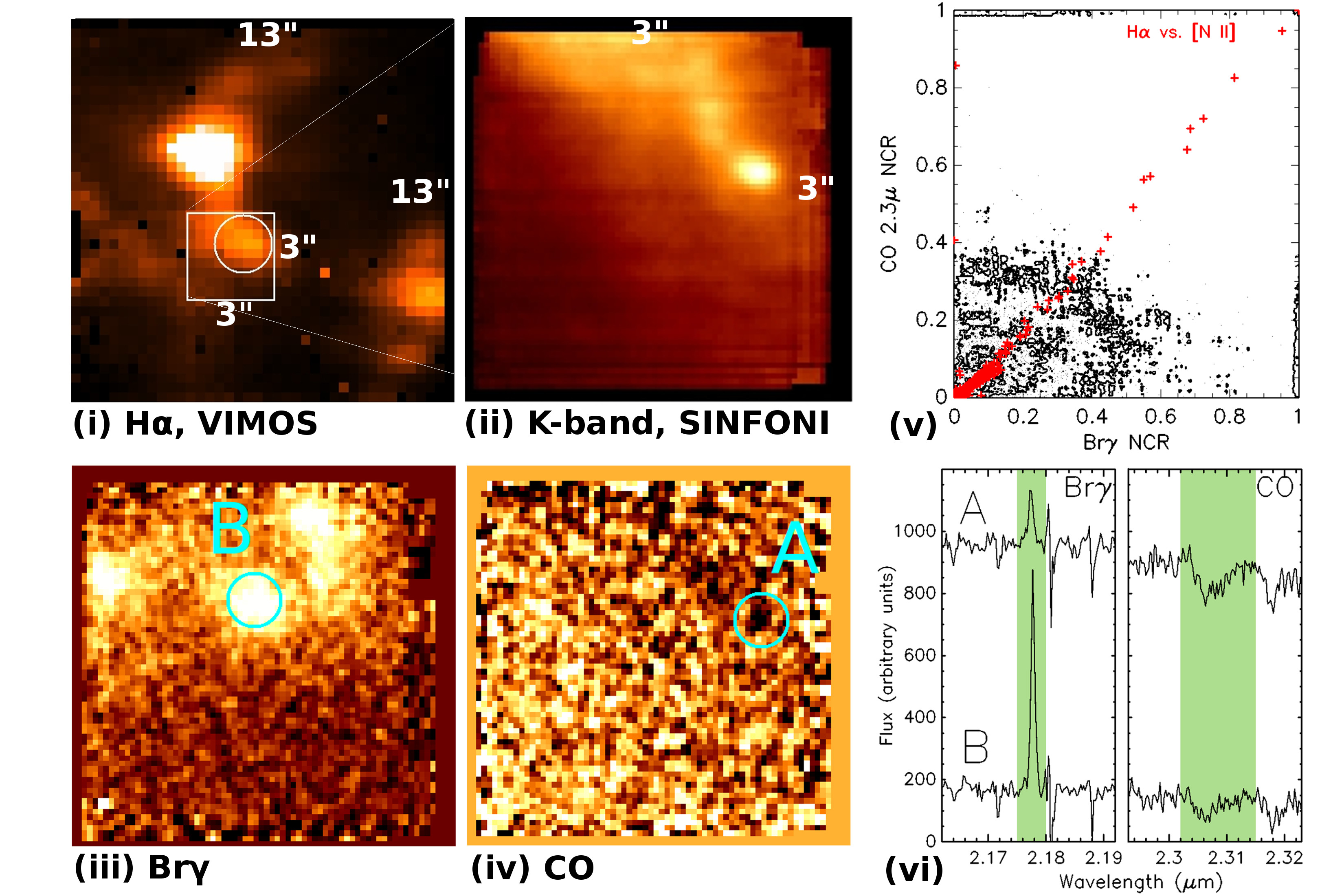}
\caption{Maps of the explosion site of SN 2009dq generated from the datacubes, in continuum-subtracted H$\alpha$ \textit{(i)}, K-band \textit{(ii)}, Br$\gamma$~\textit{(iii)}, and CO \textit{(iv)}. As the maps are color-coded with white for high-count pixels and black for low-count pixels, the pixels with strong CO absorption appear black in panel \textit{(iv)}.
The white 3" $\times$ 3" square in panel \textit{(i)} represents the SINFONI FoV, and the white circle denotes the SN position within 1" radius. North is up and east is left in all maps. Cyan circles in panels \textit{(iii)} and \textit{(iv)} indicated with A and B have identical radii of 0.25". 
Panel \textit{(v)} shows the SINFONI spaxels plotted according to the Br$\gamma$ and CO NCRs; these are shown as contours at 10\%, 50\%, and 90\% levels, while for comparison, the VIMOS spaxels are plotted in red plus marks according to the H$\alpha$ and [N \textsc{ii}] NCR. 
Panel \textit{(vi)} shows the extracted spectra of regions A and B, in the Br$\gamma$ and CO spectral regions. The two lines are indicated with green-shaded regions. Wavelength is in the observer frame.
\label{brgco}}
\end{figure*}

Br$\gamma$ emission and CO absorption are observed in some of the SN sites in the SINFONI sample. 
In the SN sites where Br$\gamma$ and CO appear together, no spatial correlation between the two lines is observed (e.g. Fig. \ref{brgco}, panels \textit{(iii)} and \textit{(iv)}). 
A more quantitative description of this observation is obtained by analysing the pixel statistics \citep{fruchter06,james06} of Br$\gamma$ and CO. In the pixel statistics technique, the IFU spaxels are ranked in a normalised cumulative rank (NCR) in which the brightest pixel has NCR = 1 while the faintest NCR = 0. This is done for both Br$\gamma$ and CO, and additionally for the H$\alpha$ and [N \textsc{ii}] emission lines in the optical data.
In panel \textit{(v)} of Fig.~\ref{brgco}, the pixel statistics for the two NIR age indicators are plotted as contours. No correlation is observed between the two, while for comparison, the pixels containing H$\alpha$ and [N~\textsc{ii}] show high correlation. This means that the brightest H$\alpha$ pixels are also the brightest [N~\textsc{ii}] pixels, while there is no such behaviour for Br$\gamma$ emission and CO absorption observed. 

While in general there is no spatial correlation between Br$\gamma$ and CO within a few hundred pc area at the explosion sites (typical SINFONI FoV size for the objects in the sample), it is observed that these two indicators can still appear together at the same position. Comparing the spectra of regions A (strong K-band source, indicative of high stellar mass) and B (strong, localised Br$\gamma$ emission) in Fig. \ref{brgco}, panel \textit{(vi)}, it is apparent that both regions contain the Br$\gamma$ and CO lines simultaneously. 

Table \ref{tab:sinfo} lists the SN explosion sites for which we have collected NIR SINFONI data, together with the values of Br$\gamma$ and CO EWs and the corresponding ages. 
To measure the Br$\gamma$ and CO EWs at the SN position, a 1-dimensional spectrum was extracted from the SINFONI datacube within one optical seeing radius. Then, Br$\gamma$ and CO lines in the spectrum were measured in the same fashion as the optical lines, and similarly the EW values were compared to Starburst99 SSP models to obtain the age estimate.
In the last column of Table \ref{tab:sinfo}, the stellar population age derived from H$\alpha$EW is  presented for comparison. The ages derived from Br$\gamma$ and H$\alpha$ are consistent with each other within $\sim15 \%$, as expected if these line emissions come from the same component in the stellar population.

In the SN sites where both Br$\gamma$ and CO are detected, the corresponding ages show that these two age indicators indeed do not probe the same age range. 
It is interesting to note that the derived CO ages, within the errors, correspond to the age range that give rise to RSG progenitors. On the other hand, the presence of Br$\gamma$ emission suggests that relatively more massive stars are also existent at the same location.
Therefore these instances illustrate the case where there could be multiple star formation bursts in a SN explosion site. 
While the youngest starbursts generally occur in compact, localized regions (Br$\gamma$, Fig. \ref{brgco} panel \textit{(iii)}), the older populations are more spread in the region (CO, Fig. \ref{brgco} panel \textit{(iv)}), forming a background. Therefore, the old population may contaminate the young population in the same line of sight.
As the SN exploded in this kind of region, the non-instantaneous star formation history may confuse the age determination. The H$\alpha$/Br$\gamma$EW method that is used in this work is sensitive towards the youngest stellar populations at the massive end of the IMF; in this context contamination from older stellar populations could be considerable and as a result the H$\alpha$/Br$\gamma$EW age is to be treated as effectively a lower limit, i.e. upper limit for the progenitor mass.

\begin{table*}
\caption{SN sites observed with SINFONI. 
} 
\label{tab:sinfo}      
\centering                          
\begin{tabular}{l c c c c c}        
\hline\hline                 
SN (type) & Br$\gamma$ EW ($\AA$) & CO EW ($\AA$) & Br$\gamma$EW age (Myr) & COEW age (Myr) & H$\alpha$EW age (Myr) \\    
\hline                        
   1970A (II)  &  $4.6\pm1.5$ & not detected  & $7.90_{-0.26}^{+0.55}$  & ---  & 7.18$^{+0.30}_{-0.16}$  \\
   1985P (IIP) &  not detected & not detected & --- & --- & not detected \\
  1992ba (IIP) & not detected & not detected & --- & --- & $6.29^{+0.13}_{-0.06}$ \\
   1997X (Ic) & $3.2\pm2.7$ & $-13.4\pm5.8$ & $7.28^{+2.14}_{-0.51}$ & $> 7.54$ &  $6.32^{+0.03}_{-0.04}$ \\
   1999br (IIP) &  not detected & not detected & --- & --- &  $6.88^{+0.10}_{-0.17}$\\ 
   2000ew (Ic) &  not detected & not detected & --- & --- & $5.75^{+0.09}_{-0.09}$ \\
   2004dg (IIP) + 2012P (IIb) & $91.2\pm5.2$ & not detected & $5.21^{+0.13}_{-0.04}$ &--- & $5.85^{+0.07}_{-0.07}$ \\     
   2009dq (IIb) & $7.9\pm0.7$ & $-12.9\pm2.3$ & $6.66^{+0.03}_{-0.03}$ & $> 7.70$ & $6.24^{+0.05}_{-0.05}$ \\
   2012A (IIP) &  not detected & not detected & --- & --- & $6.52^{+0.21}_{-0.17}$ \\
   2012au (Ib) & not detected & not detected & --- & --- & $6.02^{+0.09}_{-0.06}$ \\
   1923A (II) & $8.2\pm3.0$ & $-16.1\pm3.8$ & $6.65^{+0.20}_{-0.11}$ & $> 7.85$ & --- \\
\hline                                   
\end{tabular}
\tablefoot{
CO EW is presented with negative value to indicate that it is measured from absorption line.
SN 1923A explosion site at NGC 5236 was observed with SINFONI, but not included in the main optical sample for analysis. SNe~2004dg and 2012P occurred very close together within $\sim2$", on the same H \textsc{ii} region.
The flag "not detected" indicates that the line is not observed at the SN position, within $ 3 \sigma$ detection.
}
\end{table*}


\begin{figure*}
\centering
\begin{subfigure}{.5\textwidth}
  \centering
  \includegraphics[width=\linewidth]{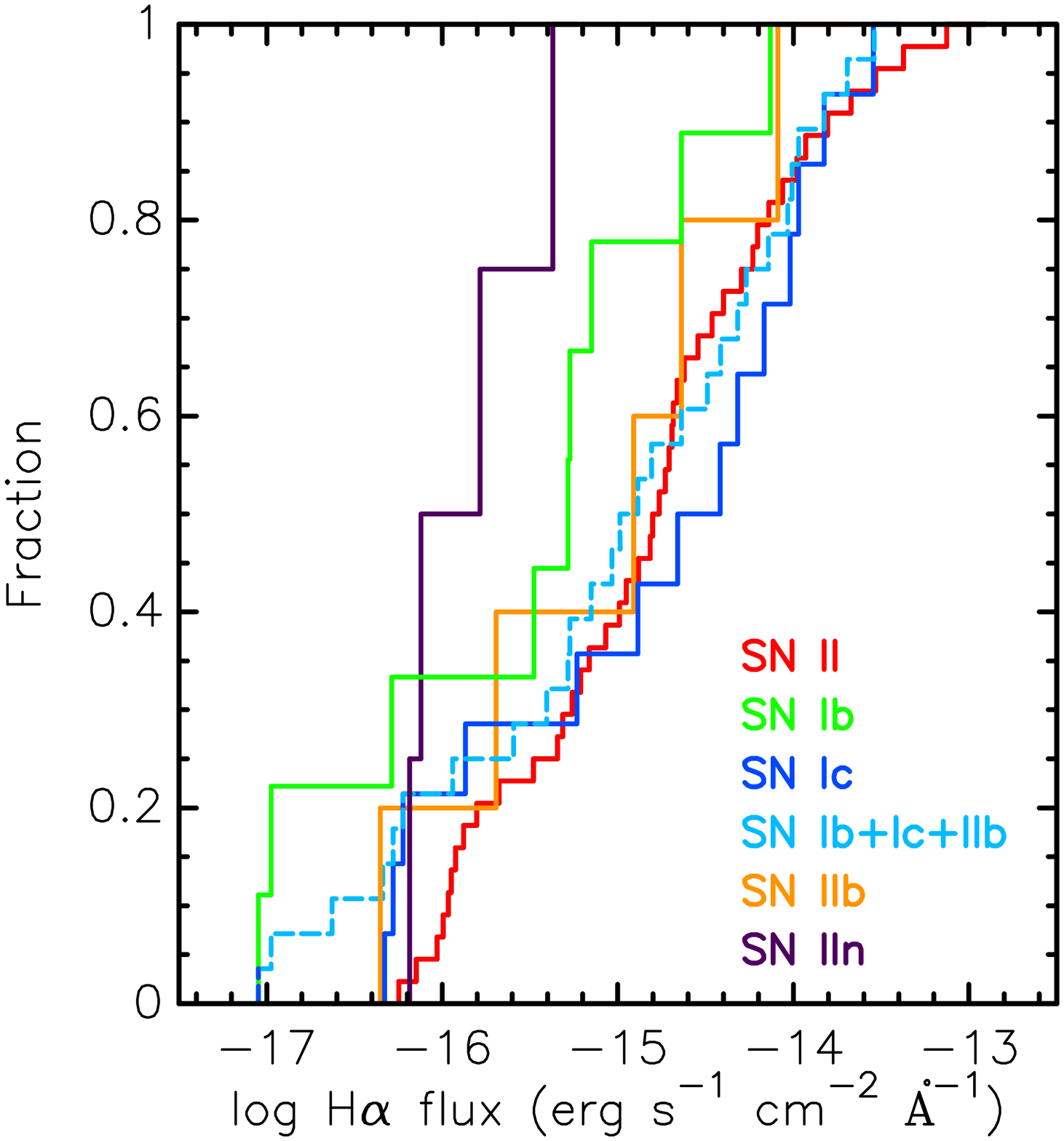}
\end{subfigure}%
\begin{subfigure}{.5\textwidth}
  \centering
  \includegraphics[width=\linewidth]{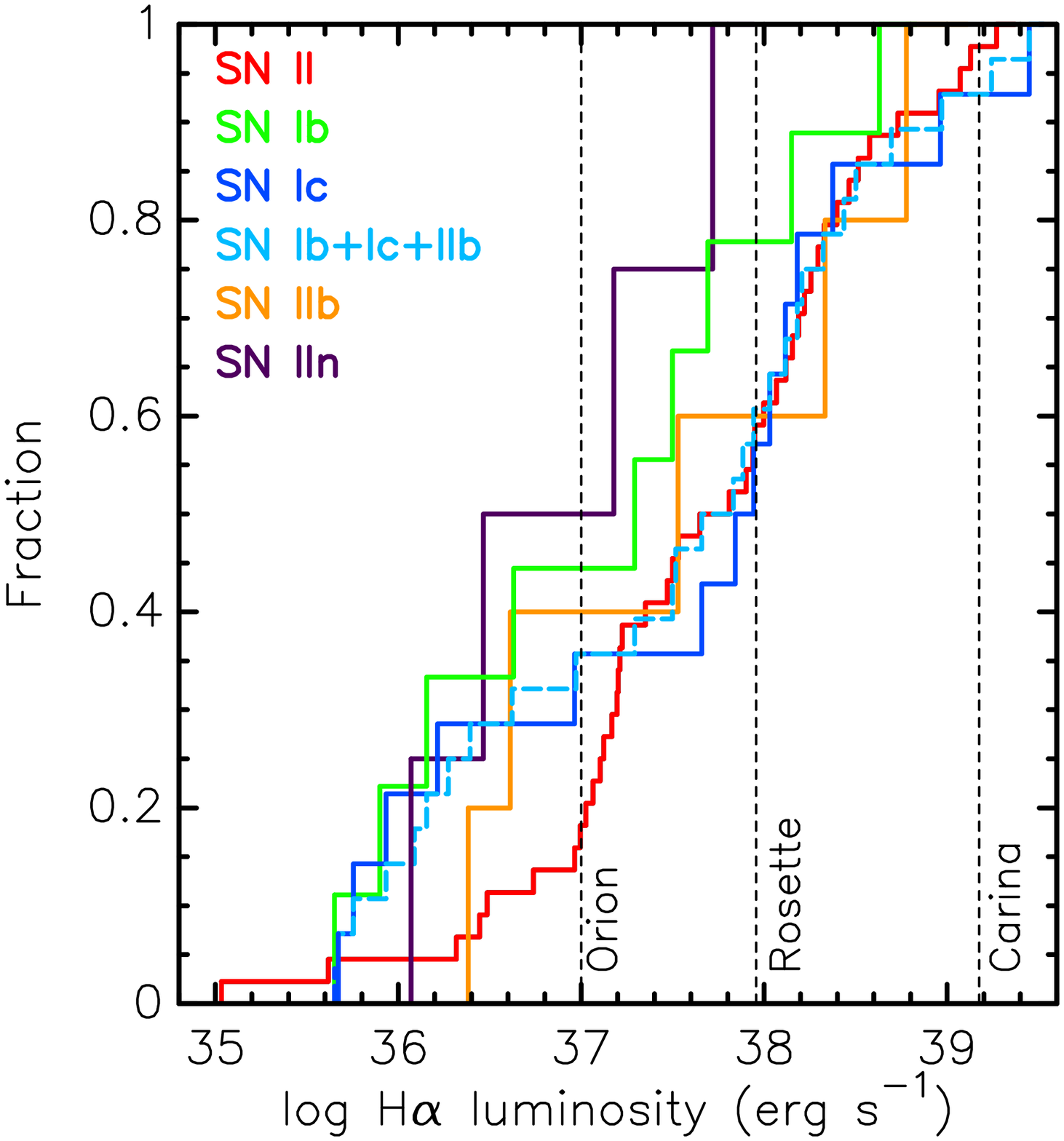}
\end{subfigure}
\caption{Cumulative distribution of host H II region flux and luminosity.
\label{edf_h2reg}}
\end{figure*}

Clues on the star formation history can also be obtained from the host H~\textsc{ii} region itself.
Larger H~\textsc{ii} regions tend to contain multiple stellar populations resulting from several past starbursts, while smaller, more compact H~\textsc{ii} regions tend to have a near-instantaneous starburst \citep{crowther13}. These giant H~\textsc{ii} regions typically have luminosities on the order of log~$L(\textrm{H}\alpha) \sim 39$~erg~s$^{-1}$ and size of few hundred pc.
We examine the H~\textsc{ii} regions at the SN sites in our sample to identify which type of H~\textsc{ii} region (hence star formation history) dominates our sample. Fig.~\ref{edf_h2reg} shows the cumulative distribution of the observed flux and luminosity of the H~\textsc{ii} regions. Most (90\%) of the H~\textsc{ii} regions have luminosity lower than log~$L(\textrm{H}\alpha) = 39$~erg~s$^{-1}$. This fact suggests that our SN sites are dominated by relatively small H~\textsc{ii} regions, expected to be characterised by closer-to-instantaneous star formation history. 
Nevertheless, even though efforts have been made to ensure that we are probing small regions, we remind the reader that the effects of chance superposition between the SN and H \textsc{ii} region may always be present in this kind of environment study. Monte Carlo simulations predict that this could affect up to $\sim50 \%$ of the observed association \citep{hk13b}. 
As seen in the NIR IFU data, in three out of five SN sites in which a young population is detected via Br$\gamma$ emission, a potentially older population is also detected via CO absorption.
The risk of chance superposition increases with lower mass progenitors, which are older (longer-living) hence have more time to drift away from the actual star formation birthplace.
In addition, studies of resolved stellar populations around type-IIP SNe show that there are cases where relatively low-mass progenitor candidates ($\sim$8 $M_\odot$) coexist with higher-mass stars (15-60 $M_\odot$) within 100 pc from the SN position \citep{maund17}. Only when an appropriate age component can be identified from the mixed stellar population one can tightly constrain the SN progenitor initial mass. 
While this caveat introduces more scatter in the statistics when comparing SNe of different types, general trends of differences between different SN types are still expected to be observed.

\section{Discussion}
\label{sec:discu}

Massive stars are thought to end their lives as SNe, after stellar evolutionary processes have guided them from their ZAMS phase through to the pre-SN stage.
The initial birth mass and metallicity are considered to be the most fundamental physical parameters in driving stellar evolution \citep[see e.g.][]{heger03}. 
As the strength of mass loss via stellar wind scales with stellar mass and metallicity, one would expect the most highly stripped SNe to be associated to progenitor stars with the highest mass and metallicity, in comparison to the other SN subtypes.

As shown in Fig.~\ref{edf_z}, the differences in metallicity between different SN subclasses are not significant. This is in contradiction with what is expected from single-star evolution theory, where metallicity-driven winds are crucial: type-Ic SNe, which are the most highly stripped, would show the highest metallicity, followed by type-Ib and finally the H-rich type-II SNe. The observation, on the other hand, reveals that this is not the case. Some SNe Ic are even located in the low metallicity part of the distribution in the current sample. This result strengthens the notion that metallicity may not play an important role in deciding the resulting SN type, in accordance with other works based on SN environments \citep{anderson10,anderson15,leloudas11,galbany16}. The environments of broad-lined SNe IcBL are found to be relatively metal poor compared to the normal CCSNe, in agreement with previous studies \citep{modjaz11,galbany16}. However, we note that there are only two such SNe in the current sample.
The explosion site of SN 1998bw (the first SN to be associated with a GRB: 980425; \citealt{galama98,kruehler17}) in this study shows a relatively lower metallicity of 12+log(O/H) = 8.30 dex compared to the GRB-less SN 2009bb \citep{pignata11}, 12+log(O/H) = 8.49 dex. \citet{levesque10}, using slit spectroscopy of the explosion site concluded that the high metallicity of SN 2009bb site is consistent with typical GRB-less SNe IcBL and not with GRB hosts. Their metallicity value recalculated on the \citet{marino13} N2 scale is 12+log(O/H) = 8.52 dex. These two different cases illustrate the importance of metallicity in deciding whether a SN IcBL progenitor would also produce GRB or not \citep{modjaz08,levesque10b}. 
Progenitors with higher metallicity are not be able to spin fast enough and produce high angular momentum essential for GRB jet production, eventually producing a GRB-less SN IcBL \citep{woosley06}.

\begin{figure}
   \centering
   \includegraphics[width=\hsize]{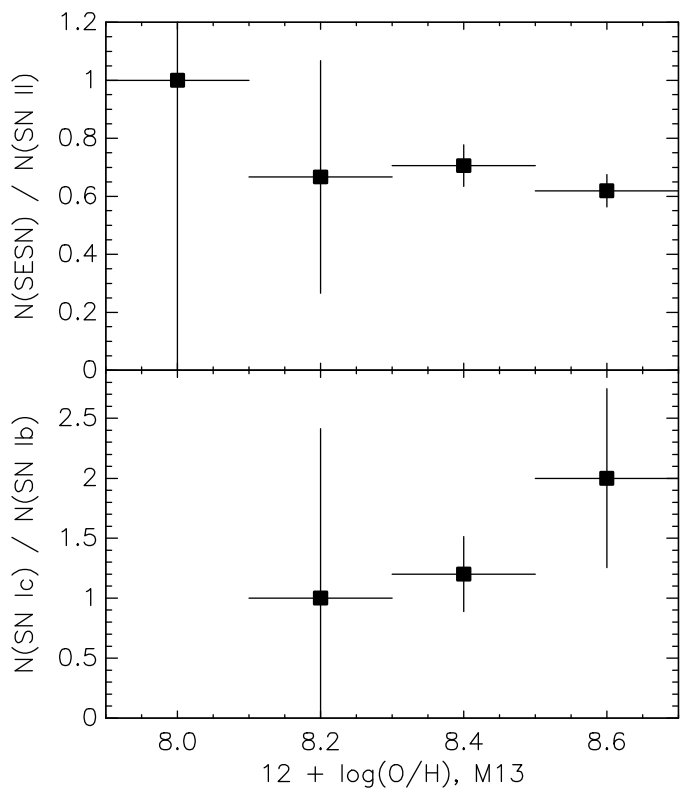}
      \caption{Number ratios of different SN types in metallicity bins: N(SESN)/N(SN II) \textit{(upper panel)}, and N(Ic)/N(Ib) \textit{(lower panel)}. Horizontal error bars indicate the bin size and vertical error bars are Poisson errors.}
         \label{z_bin}
   \end{figure}

The strong dependence of stellar wind with metallicity in massive stars should also be reflected in the number ratio of H-poor and H-rich SNe within different metallicity bins. With high metallicity, wind and mass loss become stronger, resulting in more H-poor massive stars and eventually SESNe. Fig.~\ref{z_bin} (upper panel) shows the observed number ratio between H-poor SESNe and H-rich SNe II. In general, the trend over metallicity is flat. This suggests that other factors more dominant than metallicity could be at play in producing SESNe. 
The picture changes when one examines the SN Ic/SN Ib number ratio (Fig.~\ref{z_bin}, lower panel). As metallicity increases, a rise of Ic/Ib number ratio is seen. This suggests that within the SESN group, metallicity is affecting the production of SNe Ic more than it does Ib. 
Indeed, in our sample SNe Ic are found with higher median metallicity compared to SNe Ib.
A similar trend is also found by \citet{galbany16}, where the association of SN Ic with high metallicity is even more pronounced in non-targeted surveys.
More SNe Ic are produced in high-metallicity environments, indicating that metallicity-dependent stellar winds are indeed affecting the production of this particular subclass.
This indirect evidence leaves open possibility that massive single stars with strong winds are contributing in the production of SN Ic, at least more strongly compared to SN Ib and possibly also IIb. 
However, note that binary progenitor models also predict that metallicity to some extent affect the SESN production, in a way that higher metallicity would still produce more highly-stripped progenitors \citep{yoon17}.
A standard binary model typically leaves a thin H layer after the Roche-lobe overflow, and as it is difficult to remove the He layer this way, a final push by metallicity-driven wind is suggested to be important in expelling the remaining layer and produce a SN~Ic progenitor.

A more stringent constraint for progenitor mass comes from the age analysis.
The observed H$\alpha$EW and the derived age/initial mass points toward relatively higher mass progenitors for SNe~Ic. SNe Ib and IIb, while similarly deprived of the outer hydrogen envelope, correspond closer to the lower-mass SN II progenitors \citep[see also][]{kangas17}. Again this suggests that these two He-rich subclasses are probably dominated by sub-WR mass progenitors that lost the outer envelope through binary interactions. However, statistically these results are not significant due to the small differences.

Evidence of binary interaction affecting the production of SESNe is mounting. \citet{smith11} argued that the observed relative fractions of CCSN subtypes cannot be reconciled with IMF calculations if only single WR star progenitors are expected for SESNe. There are simply not enough massive stars above 25 $M_\odot$ to account for all the observed SESNe.
\citet{lyman16} showed that the explosion parameters of most SESNe do not correspond to very massive progenitors, but moderately massive stars that were likely the production of binary interaction.
While it has been recently established that the majority of massive stars experience interaction with a binary companion during their lifetime \citep{sana12}, population synthesis studies further suggest that interaction may bring stars of intermediate mass (4-8 $M_\odot$) to core-collapse as a result of mass transfer or merger \citep{zapartas17}. 
{This accounts for 14$^{+15}_{-14}$\% of all CCSNe, }
and results in a distribution tail of `delayed' CCSNe that occur in stellar populations aged more than 50 Myr where all the massive stars above $M_{\textrm{ZAMS}} \sim 8$ $M_\odot$ have already disappeared.
In this case, these SNe would appear in regions without H$\alpha$ emission. Within our sample, 20\% of the SNe occur in such non-star forming regions.
\citet{zapartas17} speculated that the various SNe from this delayed population may not form a homogeneous class, but also noted that some type-II and IIn SNe can be produced through this scenario.

It is clear that in order to explode as SESN the progenitor star should experience significant mass loss, whose mechanism may take the form of strong binary interaction or metallicity-driven winds. Type-IIn SNe, on the other hand, are evidently interacting with nearby circumstellar material (CSM) that is thought to be the result of mass loss activity of the progenitor. At least some of these SNe IIn are associated with stars of extremely high initial mass that resemble luminous blue variable (LBV) stars \citep[see e.g.][]{smith11b}. However, in many instances the environments of SNe IIn do not suggest recent star formation \citep{anderson15}. 
In comparison with other SN types it has also been shown that SNe IIn exhibit the least association with ongoing star formation compared to the other CCSNe \citep{habergham14}. 
\citet{kangas17} showed that the SN IIn population does not share a similar spatial distribution in host galaxy H$\alpha$ light with LBV stars. They are instead best matched with the RSG stars, whose distribution suggests relatively lower mass compared to LBV stars.
An opposing view against LBVs as highly massive stars comes from \citet{smith15}, who argued that most LBV stars are isolated, being a product of binary evolution, and therefore consistent with SNe IIn exploding in passive regions.
All the evidence points to the majority of SN IIn coming from the low-mass end of massive stars.
The current study confirms this view. SNe IIn are found to be the least associated with recent star formation among CCSNe. 
While the environments suggest a low-mass origin, the fact that some SNe IIn are strongly associated with massive LBV stars would indicate that the SN IIn population is a mix of different kinds of progenitor systems. The majority of SNe IIn are produced by relatively low-mass stars as opposed to massive progenitors. 
It is interesting to note that while SNe Ic and IIn must both suffer significant progenitor mass loss prior to the SN, they represent two opposite ends in CCSN progenitor initial mass distribution. 
{Nevertheless, the mass loss mechanism for SN Ic and IIn progenitors may be different. 
SN IIn progenitors must experience significant mass loss at least immediately before the core collapse, within years to centuries \citep{smith16}. This suggests that the high mass loss rate is timed with the core collapse, and may be caused by late-stage burning instabilities \citep{smitharnett14}. On the other hand, the WR star progenitors of SNe Ic lose mass through relatively steady, strong stellar winds \citep{crowther07,smith14}.
If SN IIn progenitors only experience severe mass loss immediately before the core-collapse, this could be consistent with a scenario in which low-mass stars with average mass loss rates undergo outbursts before the core-collapse.
}

{A direct link between SNe Ic and IIn has recently been discovered. SN~2017dio showed SN Ic spectral characteristics at the early phase, and signs of an associated, H-rich CSM \citep{hk17}. At later phases, the spectral characteristics became more similar to SNe IIn as the CSM interaction became more intense. This nearby, H-rich CSM cannot come from the SN progenitor star itself as it has been stripped of the H and He layers (hence the SN Ic spectral appearance). It was suggested that the SN progenitor had a secondary companion that was H-rich due to mass transfer from the primary, and undergoing an unstable LBV phase or transferring mass back to the primary at the time of the explosion. This might explain the presence of the H-rich CSM.
}
{Another} event, SN 2014C, initially showed a type-Ib spectrum around maximum light, before transforming into showing type-IIn spectra at late times \citep{milisav15}. This event was interpreted as an explosion of a H-free progenitor star inside a CSM cavity, and as the SN ejecta traverses outward it encountered the H-rich CSM and produced IIn-like spectral signatures. The CSM must have been produced by mass loss episodes of the progenitor preceding the terminal SN.
Such events are rare and highlights our incomplete understanding of pre-SN massive star evolution.

\begin{figure}
   \centering
   \includegraphics[width=\hsize]{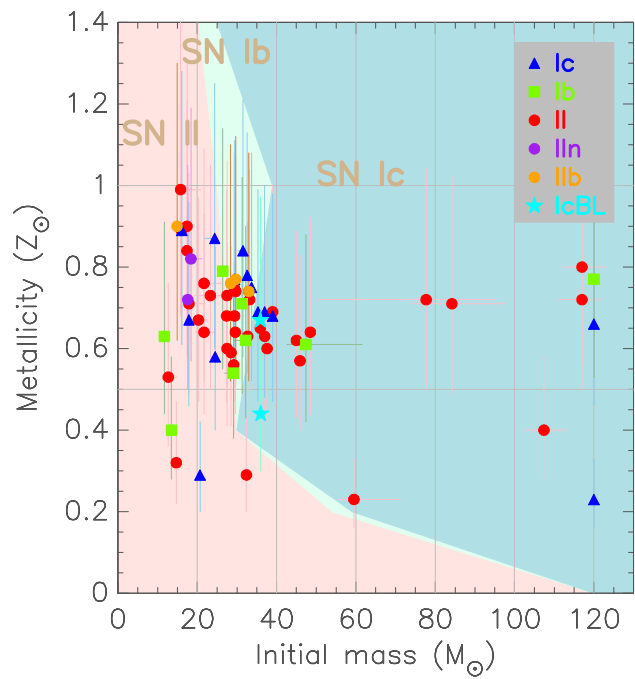}
      \caption{Diagram showing the plane of progenitor star initial mass and metallicity. Shaded areas are stellar evolution predictions from rotating single-star models of \citet{georgy09}, for SN II (pink), Ib (light green), and Ic (blue-green). Data points from this work are colour coded blue for SN Ic, bright green for Ib, red for II, purple for IIn, orange for IIb, and cyan for Ic-BL.
      Uncertainties in metallicity reflect the 0.16 dex error in the 12+log(O/H) N2 calibration of \citet{marino13}.}
         \label{mzdiag}
   \end{figure}

In the context of the whole CCSN subclass, the current study confirms that a strictly single-star progenitor scenario does not likely hold. Figure~\ref{mzdiag} shows single-star theoretical predictions of massive star evolution on the initial mass-metallicity plane \citep{georgy09}. In such a scenario, progenitors of different SN subtypes should not overlap on the mass-metallicity (M-Z) plane. However, observational data obtained in this study suggest that they are actually significantly overlapping. Most notable are the H-poor SNe that are less massive than 25 $M_\odot$ and fall into the SN~II progenitor domain. These SNe are presumably produced by progenitor stars in a binary system, since as single stars they are not massive enough to evolve into WR stars. 
Theoretical predictions that take into account binary progenitors \citep[e.g.][]{zapartas17,eldridge09} do predict that stars lower than 25 $M_\odot$ can remove the outer envelope through binary interaction. These progenitors would appear practically anywhere in the M-Z diagram (Eldridge \& Stanway, priv. comm.).

However, there still could be some influence of single progenitors in the production of CCSNe. According to the degree of envelope stripping, SESNe can be sorted as type IIb, then Ib, and Ic. Although not apparent in metallicity distributions, this pattern is visible in Figures \ref{edf_haew} and \ref{edf_m} albeit without statistical significance. 
The median initial mass estimates for SNe IIb, Ib, and Ic progenitors respectively are 29.7, 31.3, and 32.6 $M_\odot$. The two broad-lined IcBL SNe are similarly of 36 $M_\odot$, higher compared to the median of the other SN subtypes.
For comparison, SNeII progenitors have median initial mass of 29.7 $M_\odot$, similar to SNe IIb. These SN II progenitors have their hydrogen envelope largely intact at the time of the SN, and are traditionally thought to occupy the lowest mass range of CCSNe at $\lesssim 20$ $M_\odot$ through various independent methods (e.g. direct detections, \citealt{smartt09}, and nebular phase analysis, \citealt{jerkstrand12}).
With different prescriptions for mass loss and reddening, however, it is still possible to have RSG SN II progenitors at around 25 $M_\odot$ \citep{beasor16}.
It is to be noted that the initial mass estimates presented in the current work do not necessarily reflect the true value, but are more likely to be upper limits (see sections \ref{sec:age} and \ref{sec:sfh}). The nature of environment studies does not allow stringent characterization of each individual event based on the underlying stellar population, nevertheless the statistics are useful to compare and constrain different SN subclasses.

On average, SNe with a higher degree of stripping are associated with younger and more massive progenitors. This mass sequence is expected in single SN progenitor population \citep{heger03,georgy09}, and such a trend has been previously shown with SN environment observations \citep{anderson12}. Considering previous studies in the literature and the results presented in this work, the view that there is a mix between single and binary SN progenitor populations appears to be consistent with the observed signatures. Massive stellar evolution does not seem to work as a straight pipeline in which a certain fate of SN type awaits a star of a particular initial mass, but mass loss (whether induced by binary interaction or not) does affect greatly the outcome of the evolution \citep[c.f.][]{smith14}.





\section{Summary and conclusions}
\label{sec:conc}

In this work we present progenitor initial mass and metallicity constraints for distinct subclasses of CCSNe, namely SNe type II, IIn, IIb, Ib, Ic, and IcBL. As in \citet{hk13a,hk13b}, the parent stellar population of the SN was identified and analysed using IFU spectroscopy, where the integrated spectrum was used to derive the metallicity and age. 
Assuming coevality between the SN progenitor and stellar population, these two parameters were adopted for the SN progenitor and used for the derivation of its initial mass estimate. 
Unlike in \citet{hk13a,hk13b}, where the SN sites were preferentially selected for bright H \textsc{ii} regions, the current work uses a relatively unbiased, distance-limited approach while still keeping the sample distance short. 
Ninety per cent of the sample falls within $\sim$30~Mpc distance, with median sample distance of 18~Mpc and typical projected linear size per spatial resolution better than 100~pc.

The following conclusions were derived from the analysis:
\begin{itemize}
\item Metallicity differences between SN types are not statistically significant. The implication is that metallicity does not play a critical role in pre-SN mass loss and deciding the outcome SN. Nonetheless there is a subtle effect of metallicity in the production of SNe Ib and Ic. The Ic/Ib number ratio tend to increase as metallicity increases. In comparison, the number ratio between SESN and SN II across metallicity is flat.
\item SNe Ic appear to be the most associated with the youngest stellar populations and most massive progenitors. They are followed consecutively by SNe Ib, then IIb and II. SN IIn is at the opposite end of the spectrum, being associated with older populations and less massive progenitors. The differences in progenitor mass estimates are not significant, except when comparing SN IIn with other SN types.
On average, the two SNe IcBL appear to have more massive progenitors than the other CCSN progenitors. 
\item Near-IR IFS shows that there could be multiple stellar populations with different ages present at the SN site. As H$\alpha$ probes the youngest stellar populations, mass constraints derived from it should be treated as an upper limit.
\item On the initial mass-metallicity plane, SN progenitors do not conform to single-star stellar evolution predictions. Combined with evidence derived from other methods, this confirms that the progenitors of CCSNe cannot come from single stars only. There must be a significant contribution from massive interacting binaries in CCSN production.
\item Along with initial mass, mass loss is one of the most important parameters in driving massive star evolution and deciding the endpoint SN. This is apparent when comparing SNe Ic and IIn, where the progenitors of either types must similarly undergo significant mass loss and yet they populate the two opposing ends of the mass spectrum.
\end{itemize}

\begin{acknowledgements}
We thank the referee for useful suggestions that improved the paper, and Thomas Kr\"{u}hler for his indispensable help on working with MUSE.
HK acknowledges support provided by CONICYT through FONDECYT grant 3140563, and  the Ministry of Economy, Development, and Tourism's Millennium Science Initiative through grant IC120009, awarded to The Millennium Institute of Astrophysics, MAS.  
KM acknowledges support provided by Japan Society for the Promotion of Science (JSPS) through KAKENHI Grant 17H02864 and through JSPS Open Partnership Bilateral Joint Research Project between Japan and Chile.
LG and KM acknowledge support from FINCA visitor program.
Based on observations collected at the European Organisation for Astronomical Research in the Southern Hemisphere under ESO programmes 089.D-0367, 091.D-0482, 093.D-0318, 094.D-0290, and 095.D-0172.
It is a pleasure to thank K. Nomoto, M.~Tanaka, S.~Mattila and T. Kangas for fruitful discussions, J. Lyman for alerting us to the discovery of SN 2017ahn, and G.~Pignata for sharing the FITS image of SN 2009bb.
HK is indebted to the late Dading H. Nugroho for his suggestions in working with datacubes.

\end{acknowledgements}

%
%

\begin{appendix}
\section{Long tables}

\longtab[1]{
\begin{longtable}{lccccccr}
\caption{\label{tab:obj} Object list}\\
\hline\hline
SN & Type  & Host galaxy & $d$ (Mpc)\tablefootmark{1} & Obs. date\tablefootmark{2} & Instrument & Seeing\tablefootmark{3} & Resolution (pc)\tablefootmark{4} \\
\hline
\endfirsthead
\caption{continued.}\\
\hline\hline
SN & Type  & Host galaxy & $d$ (Mpc)\tablefootmark{1} & Obs. date\tablefootmark{2} & Instrument & Seeing\tablefootmark{3} & Resolution (pc)\tablefootmark{4} \\
\hline
\endhead
\hline
\endfoot
1970A	&	  II:	&	  IC 3476	& $	13.3	\pm	0.2	$ &	2015 May 14 (opt)	&	MUSE/SINFONI	&	1”.0	&	64.5	\\
1970G*	&	  IIL	&	  NGC 5457	& $	6.95	\pm	0.06	$ &	2011 Mar 10	&	SNIFS	&	1“.3	&	43.8	\\
1978G	&	  IIn	&	  IC 5201	& $	10.57	\pm	0.2	$ &	2015 May 15	&	MUSE 	&	1”.0	&	51.2	\\
1978K	&	  IIn?	&	  NGC 1313	& $	4.25	\pm	0.08	$ &	2014 Nov 24	&	VIMOS 	&	0”.8	&	16.5	\\
1982R	&	  Ib	&	  NGC 1187	& $	22.18	\pm	0.2	$ &	2014 Nov 23	&	VIMOS 	&	0”.7	&	75.3	\\
1983K	&	  IIP	&	  NGC 4699	& $	15.3	\pm	1	$ &	2015 May 14	&	MUSE 	&	0”.6	&	44.5	\\
1983N*	&	  Ib	&	  NGC 5236	& $	4.66	\pm	0.07	$ &	2011 Mar 11	&	SNIFS	&	1”.5	&	33.9	\\
1984E	&	  IIL	&	  E184-G82	& $	23.23	\pm	0.1	$ &	2014 Apr 3	&	VIMOS 	&	1”.0	&	112.6	\\
1984L*	&	  Ib	&	  NGC  991	& $	17.3	\pm	1.1	$ &	2011 Sep 29	&	GMOS-N	&	0”.6	&	50.3	\\
1985G	&	  IIP	&	  NGC 4451	& $	25.94	\pm	0.2	$ &	2015 May 16	&	MUSE 	&	0”.8	&	100.6	\\
1985P	&	  IIP	&	  E184-G82	& $	16.8	\pm	1	$ &	2014 Nov 23 (opt)	&	VIMOS/SINFONI	&	0”.7	&	57.0	\\
1987K	&	  IIb:	&	  NGC 4651	& $	29.11	\pm	0.2	$ &	2015 May 15	&	MUSE 	&	0”.8	&	112.9	\\
1988E	&	II	&	  NGC 4772	& $	15.6	\pm	1	$ &	2015 May 17	&	MUSE 	&	0”.8	&	60.5	\\
1990Q	&	  II	&	  NGC 5917	& $	28	\pm	5.5	$ &	2015 May 14	&	MUSE 	&	0”.5	&	67.9	\\
1992ba	&	  IIP	&	  NGC 2082	& $	13.1	\pm	1.8	$ &	2014 Nov 24 (opt)	&	VIMOS/SINFONI	&	0”.8	&	50.8	\\
1992bd	&	  II	&	  NGC 1097	& $	16	\pm	0.2	$ &	2014 Nov 24	&	VIMOS 	&	0”.8	&	62.1	\\
1994I*	&	  Ic	&	  NGC 5194	& $	8.39	\pm	0.84	$ &	2010 Mar 10	&	SNIFS	&	1”.1	&	44.7	\\
1994L*	&	  II	&	  NGC 2848	& $	20.61	\pm	0.2	$ &	2011 Mar 11	&	SNIFS	&	1”.3	&	129.9	\\
1996an	&	  II	&	  NGC 1084	& $	20.89	\pm	0.2	$ &	2014 Nov 24	&	VIMOS 	&	0”.9	&	91.1	\\
1996N	&	  Ib	&	  NGC 1398	& $	22.7	\pm	1.3	$ &	2014 Nov 24	&	VIMOS 	&	0”.8	&	88.0	\\
1997dn	&	  II	&	  NGC 3451	& $	28.18	\pm	0.2	$ &	2015 May 14	&	MUSE 	&	0”.6	&	82.0	\\
1997dq	&	  Ic	&	  NGC 3810	& $	16.37	\pm	0.2	$ &	2014 Apr 3	&	VIMOS 	&	0”.7	&	55.6	\\
1997X	&	  Ic	&	  NGC 4691	& $	12	\pm	0	$ &	2015 May 17 (opt)	&	MUSE/SINFONI	&	1”.0	&	58.2	\\
1998bw	&	  Ic	&	  E184-G82	& $	37.9	\pm	2.7\dagger	$ &	2015 May 14/15	&	MUSE 	&	1”.0	&	192.5	\\
1998dl	&	  IIP	&	  NGC 1084	& $	20.89	\pm	0.2	$ &	2014 Nov 24	&	VIMOS 	&	0”.9	&	91.1	\\
1998dn	&	  II	&	  NGC 337A	& $	11.4	\pm	2.1	$ &	2014 Jul 21	&	VIMOS 	&	0”.9	&	49.7	\\
1999br	&	  IIP 	&	  NGC 4900	& $	15.6	\pm	1	$ &	2015 May 15 (opt)	&	MUSE/SINFONI	&	0”.8	&	60.5	\\
1999ec*	&	  Ib	&	  NGC 2207	& $	31.6	\pm	2.6	$ &	2011 Mar 10	&	SNIFS	&	1”.1	&	168.5	\\
1999eu	&	  IIP 	&	  NGC 1097	& $	16	\pm	0.2	$ &	2014 Nov 24	&	VIMOS 	&	0”.8	&	62.1	\\
1999gi*	&	  IIP	&	  NGC 3184	& $	13	\pm	0	$ &	2011 Mar 11	&	SNIFS	&	1”.0	&	63.0	\\
1999gn*	&	  IIP	&	  NGC 4303	& $	17.6	\pm	0.9	$ &	2011 Mar 15	&	SNIFS	&	0”.8	&	68.3	\\
2000cl	&	  IIn	&	  NGC 3318	& $	35.32	\pm	0.12	$ &	2015 May 17	&	MUSE 	&	0”.8	&	137.0	\\
2000ew*	&	  Ic	&	  NGC 3810	& $	16.37	\pm	0.2	$ &	2011 Mar 15	&	SNIFS/SINFONI	&	0”.9	&	71.4	\\
2000P	&	  IIn	&	  NGC 4965	& $	30.5	\pm	3.1	$ &	2015 May 17	&	MUSE 	&	0”.6	&	88.7	\\
2001fv	&	  IIP	&	  NGC 3512	& $	26.1	\pm	2	$ &	2015 May 16	&	MUSE 	&	0”.8	&	101.2	\\
2001ig	&	  IIb	&	  NGC 7424	& $	7.94	\pm	0.77	$ &	2014 Jul 21	&	VIMOS 	&	1”.0	&	38.5	\\
2001X	&	  IIP	&	  NGC 5921	& $	14	\pm	3.2	$ &	2015 May 15	&	MUSE 	&	0”.7	&	47.5	\\
2002hh*	&	  IIP	&	  NGC 6946	& $	7	\pm	0	$ &	2010 Aug 1	&	SNIFS	&	0”.8	&	27.1	\\
2003B	&	  IIP	&	  NGC 1097	& $	16	\pm	0.2	$ &	2014 Nov 23	&	VIMOS 	&	0”.7	&	54.3	\\
2003gf	&	  Ic:	&	  M-04-52-26	& $	37.4	\pm	2.6\dagger	$ &	2014 Jul 21	&	VIMOS 	&	1”.0	&	181.3	\\
2003ie*	&	  IIP	&	  NGC 4051	& $	11.02	\pm	0.2	$ &	2011 Mar 13	&	SNIFS	&	0”.8	&	42.7	\\
2003jg	&	  Ib/c	&	  NGC 2997	& $	11.3	\pm	0.8	$ &	2014 Apr 3	&	VIMOS 	&	0”.9	&	49.3	\\
2004am*	&	  IIP	&	  NGC 3034	& $	3.53	\pm	0.08	$ &	2011 Mar 10	&	SNIFS	&	1”.1	&	18.8	\\
2004ao	&	  Ib	&	  UGC 10862	& $	25.9	\pm	4.7	$ &	2014 Jul 21	&	VIMOS 	&	1”.2	&	150.7	\\
2004dg	&	  IIP	&	  NGC 5806	& $	26.79	\pm	0.2	$ &	2014 Apr 4 (opt)	&	VIMOS/SINFONI	&	0”.8	&	103.9	\\
2004dj*	&	  IIP	&	  NGC 2403	& $	3.18	\pm	0.06	$ &	2011 Mar 10	&	SNIFS	&	1“.3	&	20.0	\\
2004gt*	&	  Ic	&	  NGC 4038	& $	22.08	\pm	0.1	$ &	2011 Mar 10	&	SNIFS	&	1”.2	&	128.5	\\
2005at	&	  Ic	&	  NGC 6744	& $	9.15	\pm	0.09	$ &	2015 May 14	&	MUSE 	&	0”.7	&	31.1	\\
2005ay*	&	  IIP	&	  NGC 3938	& $	17.1	\pm	0.8	$ &	2011 Mar 15	&	SNIFS	&	1”.4	&	116.1	\\
2006ca	&	  II	&	  UGC 11214	& $	38	\pm	0	$ &	2014 Jul 21	&	VIMOS 	&	1”.2	&	221.1	\\
2006my	&	  IIP	&	  NGC 4651	& $	29.11	\pm	0.2	$ &	2015 May 15	&	MUSE 	&	0”.8	&	112.9	\\
2007aa	&	  IIP	&	  NGC 4030	& $	29.92	\pm	0.2	$ &	2015 Feb 13	&	VIMOS 	&	0”.8	&	116.0	\\
2007gr*	&	  Ic	&	  NGC 1058	& $	9.86	\pm	0.61	$ &	2011 Sep 29	&	GMOS-N	&	0”.5	&	23.9	\\
2007it	&	  II	&	  NGC 5530	& $	12.3	\pm	0.2	$ &	2015 Feb 13	&	VIMOS 	&	0”.9	&	53.7	\\
2007oc	&	  IIP	&	  NGC 7418A	& $	23.7	\pm	1.8	$ &	2014 Jul 21	&	VIMOS 	&	1”.0	&	114.9	\\
2007Y	&	  Ib	&	  NGC 1187	& $	22.18	\pm	0.2	$ &	2014 Nov 23	&	VIMOS 	&	0”.7	&	75.3	\\
2008bk*	&	  IIP	&	  NGC 7793	& $	3.58	\pm	0.07	$ &	2010 Aug 1	&	SNIFS	&	0”.8	&	13.9	\\
2008bo*	&	  Ib	&	  NGC 6643	& $	21.28	\pm	0.2	$ &	2010 Aug 1	&	SNIFS	&	0”.8	&	82.5	\\
2009bb	&	  Ic	&	  NGC 3278	& $	38.55	\pm	0.2	$ &	2015 May 15	&	MUSE 	&	0”.7	&	130.8	\\
2009dq	&	  IIb	&	  IC 2554	& $	22.9	\pm	1.9	$ &	2014 Nov 23 (opt)	&	VIMOS/SINFONI	&	0”.8	&	88.8	\\
2009em*	&	  Ic	&	  NGC 157	& $	12.08	\pm	0.2	$ &	2011 Sep 29	&	GMOS-N	&	0”.6	&	35.1	\\
2009H	&	  II	&	  NGC 1084	& $	20.89	\pm	0.2	$ &	2014 Nov 23	&	VIMOS 	&	0”.9	&	91.1	\\
2009hd*	&	  II	&	  NGC 3627	& $	9.04	\pm	0.07	$ &	2011 Mar 15	&	SNIFS	&	0”.6	&	26.3	\\
2009ib	&	  IIP	&	  NGC 1559	& $	12.59	\pm	0.2	$ &	2015 Feb 13	&	VIMOS 	&	1”.2	&	73.2	\\
2009ip	&	  IIn?	&	  NGC 7259	& $	25.8	\pm	1.8\dagger	$ &	2014 Apr 3	&	VIMOS 	&	1”.0	&	125.1	\\
2009jf*	&	  Ib	&	  NGC 7479	& $	36.8	\pm	0.2	$ &	2011 Sep 29	&	GMOS-N	&	0”.5	&	89.2	\\
2009kr*	&	  IIL	&	  NGC 1832	& $	22.28	\pm	0.2	$ &	2011 Mar 11	&	SNIFS	&	1“.3	&	140.4	\\
2009ls	&	  II	&	  NGC 3423	& $	17	\pm	2.5	$ &	2015 Feb 13	&	VIMOS 	&	1”.2	&	98.9	\\
2009md	&	  IIP	&	  NGC 3389	& $	20.8	\pm	0.2	$ &	2015 May 14	&	MUSE 	&	0”.6	&	60.5	\\
2009N	&	  IIP	&	  NGC 4487	& $	11	\pm	0.8	$ &	2015 May 17	&	MUSE 	&	0”.9	&	48.0	\\
2010F	&	  IIP	&	  NGC 3120	& $	23.88	\pm	0.2	$ &	2015 May 16	&	MUSE 	&	0”.8	&	92.6	\\
2011fh	&	  IIn	&	  NGC 4806	& $	29	\pm	0	$ &	2015 May 16	&	MUSE 	&	0”.9	&	126.5	\\
2011gv	&	  IIP	&	  IC 4901	& $	21.2	\pm	2.4	$ &	2014 Jul 21	&	VIMOS 	&	1”.0	&	102.8	\\
2011jm	&	  Ic	&	  NGC 4809	& $	15.6	\pm	1	$ &	2015 May 14/15	&	MUSE 	&	0”.6	&	45.4	\\
2012A	&	  IIP	&	  NGC 3239	& $	10	\pm	0	$ &	2015 Feb 13 (opt)	&	VIMOS/SINFONI	&	1”.1	&	53.3	\\
2012au	&	  Ib	&	  NGC 4790	& $	15.3	\pm	1	$ &	2014 Apr 4 (opt)	&	VIMOS/SINFONI	&	0”.7	&	51.9	\\
2012aw	&	  IIP	&	  NGC 3351	& $	10.47	\pm	0.06	$ &	2015 Feb 13	&	VIMOS 	&	0”.8	&	40.6	\\
2012cw	&	  Ic	&	  NGC 3166	& $	23.7	\pm	1.9	$ &	2014 Apr 3	&	VIMOS 	&	1”.0	&	114.9	\\
2012ec	&	  IIP	&	  NGC 1084	& $	20.89	\pm	0.2	$ &	2014 Nov 23	&	VIMOS 	&	0”.8	&	81.0	\\
2012fh	&	  Ic	&	  NGC 3344	& $	9.82	\pm	0.1	$ &	2015 Feb 13	&	VIMOS 	&	1”.2	&	57.1	\\
2012ho	&	  IIP	&	  M-01-57-21	& $	30.76	\pm	0.2	$ &	2015 May 14	&	MUSE 	&	1”.1	&	164.0	\\
2012P	&	  IIb	&	  NGC 5806	& $	26.79	\pm	0.2	$ &	2014 Apr 4 (opt)	&	VIMOS/SINFONI	&	0”.8	&	103.9	\\
2013F	&	  Ib/c	&	  IC 5325	& $	18.7	\pm	1.6	$ &	2014 Jul 21	&	VIMOS 	&	0”.9	&	81.6	\\
2017ahn	&	II	&	NGC  3318	&$	35.32	\pm	0.12	$ &	2015 May 17	&	MUSE 	&	0”.8	&	137.0	\\
\end{longtable}
\tablefoot{
{Entries noted with asterisk (*) are from SNIFS and GMOS-N observations of \citet{hk13a,hk13b}. \\}
\tablefoottext{1}{Distance according to the Extragalactic Distance Database (EDD, \href{http://edd.ifa.hawaii.edu/}{http://edd.ifa.hawaii.edu/}), \citet{tully09}, unless noted with {a dagger ($\dagger$)}. Entries noted with {dagger} are not available in EDD, thus the distances are computed from redshift taking into account the influence of the Virgo Cluster, the Great Attractor, and the Shapley Supercluster \citep{mould00}, as given in the NASA/IPAC Extragalactic Database (NED, \href{http://ned.ipac.caltech.edu/}{http://ned.ipac.caltech.edu/}).} \\
\tablefoottext{2}{Local date when the night starts, for the optical observations.} \\
\tablefoottext{3}{In optical wavelength regime; DIMM seeing.} \\
\tablefoottext{4}{The corresponding projected spatial resolution from the seeing size, in parsec.}
}
}

\longtab[2]{
\begin{longtable}{lccrrrr}
\caption{\label{tab:res} Results}\\
\hline\hline
SN & Type  & 12+log(O/H)\tablefootmark{1} & $Z$ ($Z_\odot$)\tablefootmark{2} & H$\alpha$EW ($\AA$) & Age (Myr) & $M_0$ ($M_\odot$)\\
\hline
\endfirsthead
\caption{continued.}\\
\hline\hline
SN & Type  & 12+log(O/H)\tablefootmark{1} & $Z$ ($Z_\odot$)\tablefootmark{2} & H$\alpha$EW ($\AA$) & Age (Myr) & $M_0$ ($M_\odot$)\\
\hline
\endhead
\hline
\endfoot
1970A	&	II:	&	8.43	& $	0.59	^{+	0.26	}_{	-0.18	}$ & $	48.4	\pm	5.2	$ & $	7.18	_{	-0.16	}^{+	0.30	}$ & $	28.5	^{+	0.5	}_{	-0.9	}$ \\
1970G*	&	IIL	&	8.26	& $	0.40	^{+	0.18	}_{	-0.12	}$ & $	990.5	\pm	42.6	$ & $	3.41	_{	-0.03	}^{+	0.05	}$ & $	107.4	^{+	6.0	}_{	-5.0	}$ \\
1978G	&	IIn	&	8.42	& $	0.57	^{+	0.25	}_{	-0.18	}$ & $	. . .			$ & $	. . .					$ & $	. . .					$ \\
1978K	&	IIn?	&	. . .	& $	. . .					$ & $	. . .			$ & $	. . .					$ & $	. . .					$ \\
1982R	&	Ib	&	8.54	& $	0.77	^{+	0.34	}_{	-0.24	}$ & $	1670.0	\pm	215.1	$ & $	2.64	_{	-0.22	}^{+	0.35	}$ & $	120.0	^{		}_{		}$ \\
1983K	&	IIP	&	8.55	& $	0.78	^{+	0.35	}_{	-0.24	}$ & $	. . .			$ & $	. . .					$ & $	. . .					$ \\
1983N*	&	Ib	&	8.56	& $	0.79	^{+	0.35	}_{	-0.24	}$ & $	23.3	\pm	4.2	$ & $	7.22	_{	-0.42	}^{+	0.27	}$ & $	26.4	^{+	1.4	}_{	-0.8	}$ \\
1984E	&	IIL	&	8.52	& $	0.72	^{+	0.32	}_{	-0.22	}$ & $	553.8	\pm	83.3	$ & $	3.81	_{	-0.32	}^{+	0.82	}$ & $	77.7	^{+	17.3	}_{	-27.5	}$ \\
1984L*	&	Ib	&	8.26	& $	0.40	^{+	0.18	}_{	-0.12	}$ & $	2.9	\pm	0.5	$ & $	18.00	_{	-2.21	}^{+	2.18	}$ & $	13.5	^{+	1.1	}_{	-0.9	}$ \\
1985G	&	IIP	&	8.54	& $	0.76	^{+	0.34	}_{	-0.23	}$ & $	19.6	\pm	2.5	$ & $	8.65	_{	-1.34	}^{+	0.14	}$ & $	21.7	^{+	4.4	}_{	-0.5	}$ \\
1985P	&	IIP	&	8.64	& $	0.96	^{+	0.43	}_{	-0.30	}$ & $	. . .			$ & $	. . .					$ & $	. . .					$ \\
1987K	&	IIb	&	8.54	& $	0.76	^{+	0.34	}_{	-0.23	}$ & $	36.5	\pm	3.9	$ & $	6.62	_{	-0.07	}^{+	0.05	}$ & $	28.4	^{+	0.3	}_{	-0.1	}$ \\
1988E	&	II	&	8.64	& $	0.96	^{+	0.43	}_{	-0.30	}$ & $	. . .			$ & $	. . .					$ & $	. . .					$ \\
1990Q	&	II	&	8.41	& $	0.56	^{+	0.25	}_{	-0.17	}$ & $	61.0	\pm	7.5	$ & $	6.92	_{	-0.17	}^{+	0.10	}$ & $	29.2	^{+	0.5	}_{	-0.2	}$ \\
1992ba	&	IIP	&	8.46	& $	0.63	^{+	0.28	}_{	-0.19	}$ & $	96.4	\pm	11.9	$ & $	6.29	_{	-0.06	}^{+	0.13	}$ & $	32.7	^{+	0.5	}_{	-0.9	}$ \\
1994I*	&	Ic	&	8.48	& $	0.67	^{+	0.30	}_{	-0.21	}$ & $	13.3	\pm	1.9	$ & $	11.00	_{	-0.75	}^{+	0.25	}$ & $	17.9	^{+	0.8	}_{	-0.4	}$ \\
1994L*	&	II	&	8.42	& $	0.57	^{+	0.26	}_{	-0.18	}$ & $	351.3	\pm	59.7	$ & $	4.99	_{	-0.14	}^{+	0.10	}$ & $	45.9	^{+	2.6	}_{	-1.9	}$ \\
1996an	&	II	&	8.50	& $	0.69	^{+	0.31	}_{	-0.21	}$ & $	221.4	\pm	23.7	$ & $	5.45	_{	-0.11	}^{+	0.16	}$ & $	39.0	^{+	0.8	}_{	-1.2	}$ \\
1996N	&	Ib	&	8.45	& $	0.62	^{+	0.28	}_{	-0.19	}$ & $	89.3	\pm	31.2	$ & $	6.36	_{	-0.20	}^{+	0.60	}$ & $	32.2	^{+	1.5	}_{	-3.1	}$ \\
1997dn	&	II	&	8.49	& $	0.67	^{+	0.30	}_{	-0.21	}$ & $	18.7	\pm	2.5	$ & $	9.98	_{	-0.81	}^{+	0.56	}$ & $	20.3	^{+	2.4	}_{	-0.8	}$ \\
1997dq	&	Ic	&	8.50	& $	0.69	^{+	0.31	}_{	-0.21	}$ & $	181.4	\pm	20.5	$ & $	5.71	_{	-0.13	}^{+	0.17	}$ & $	37.0	^{+	1.0	}_{	-1.2	}$ \\
1997X	&	Ic	&	8.54	& $	0.77	^{+	0.34	}_{	-0.24	}$ & $	66.8	\pm	7.0	$ & $	6.32	_{	-0.04	}^{+	0.03	}$ & $	29.4	^{+	0.2	}_{	-0.1	}$ \\
1998bw	&	Ic	&	8.30	& $	0.44	^{+	0.20	}_{	-0.14	}$ & $	167.3	\pm	17.2	$ & $	5.84	_{	-0.15	}^{+	0.12	}$ & $	36.0	^{+	1.2	}_{	-0.8	}$ \\
1998dl	&	IIP	&	8.46	& $	0.64	^{+	0.28	}_{	-0.20	}$ & $	411.3	\pm	44.0	$ & $	4.85	_{	-0.05	}^{+	0.10	}$ & $	48.5	^{+	0.9	}_{	-1.9	}$ \\
1998dn	&	II	&	8.01	& $	0.23	^{+	0.10	}_{	-0.07	}$ & $	1034.0	\pm	118.0	$ & $	4.23	_{	-0.26	}^{+	0.24	}$ & $	59.5	^{+	11.7	}_{	-3.9	}$ \\
1999br	&	IIP	&	8.49	& $	0.68	^{+	0.30	}_{	-0.21	}$ & $	63.3	\pm	6.7	$ & $	6.88	_{	-0.17	}^{+	0.10	}$ & $	29.4	^{+	0.5	}_{	-0.4	}$ \\
1999ec*	&	Ib	&	8.44	& $	0.61	^{+	0.27	}_{	-0.19	}$ & $	384.5	\pm	46.1	$ & $	5.34	_{	-0.35	}^{+	0.15	}$ & $	47.4	^{+	14.2	}_{	-4.8	}$ \\
1999eu	&	IIP	&	8.51	& $	0.72	^{+	0.32	}_{	-0.22	}$ & $	. . .			$ & $	. . .					$ & $	. . .					$ \\
1999gi*	&	IIP	&	8.47	& $	0.64	^{+	0.29	}_{	-0.20	}$ & $	66.9	\pm	28.8	$ & $	6.32	_{	-0.17	}^{+	0.27	}$ & $	29.6	^{+	0.5	}_{	-0.9	}$ \\
1999gn*	&	IIP	&	8.56	& $	0.80	^{+	0.36	}_{	-0.25	}$ & $	988.3	\pm	89.0	$ & $	3.26	_{	-0.05	}^{+	0.04	}$ & $	117.0	^{+	6.0	}_{	-5.0	}$ \\
2000cl	&	IIn	&	8.57	& $	0.82	^{+	0.37	}_{	-0.25	}$ & $	14.9	\pm	1.8	$ & $	10.52	_{	-1.74	}^{+	0.49	}$ & $	18.4	^{+	2.9	}_{	-0.6	}$ \\
2000ew*	&	Ic	&	8.49	& $	0.68	^{+	0.30	}_{	-0.21	}$ & $	221.1	\pm	35.4	$ & $	5.75	_{	-0.09	}^{+	0.09	}$ & $	39.0	^{+	1.0	}_{	-0.8	}$ \\
2000P	&	IIn	&	8.52	& $	0.72	^{+	0.32	}_{	-0.22	}$ & $	12.0	\pm	1.6	$ & $	11.19	_{	-0.28	}^{+	0.20	}$ & $	17.6	^{+	0.3	}_{	-0.3	}$ \\
2001fv	&	IIP	&	8.61	& $	0.90	^{+	0.40	}_{	-0.28	}$ & $	10.5	\pm	1.5	$ & $	11.37	_{	-0.19	}^{+	0.86	}$ & $	17.4	^{+	0.2	}_{	-1.1	}$ \\
2001ig	&	IIb	&	8.32	& $	0.46	^{+	0.20	}_{	-0.14	}$ & $	. . .			$ & $	. . .					$ & $	. . .					$ \\
2001X	&	IIP	&	8.58	& $	0.84	^{+	0.37	}_{	-0.26	}$ & $	10.9	\pm	1.4	$ & $	11.33	_{	-0.18	}^{+	0.82	}$ & $	17.4	^{+	0.2	}_{	-1.0	}$ \\
2002hh*	&	IIP	&	8.52	& $	0.72	^{+	0.32	}_{	-0.22	}$ & $	188.2	\pm	54.6	$ & $	5.83	_{	-0.14	}^{+	0.15	}$ & $	33.2	^{+	1.4	}_{	-1.5	}$ \\
2003B	&	Ic	&	8.55	& $	0.78	^{+	0.35	}_{	-0.24	}$ & $	164.3	\pm	22.0	$ & $	5.89	_{	-0.05	}^{+	0.06	}$ & $	32.6	^{+	0.5	}_{	-0.6	}$ \\
2003gf	&	Ic	&	8.02	& $	0.23	^{+	0.10	}_{	-0.07	}$ & $	1591.0	\pm	165.5	$ & $	2.90	_{	-0.07	}^{+	0.06	}$ & $	120.0	^{		}_{		}$ \\
2003ie*	&	IIP	&	8.52	& $	0.73	^{+	0.33	}_{	-0.23	}$ & $	. . .			$ & $	. . .					$ & $	. . .					$ \\
2003jg	&	Ib/c	&	8.50	& $	0.69	^{+	0.31	}_{	-0.21	}$ & $	151.3	\pm	19.3	$ & $	5.95	_{	-0.14	}^{+	0.14	}$ & $	35.3	^{+	1.0	}_{	-1.1	}$ \\
2004am*	&	IIP	&	8.65	& $	0.99	^{+	0.44	}_{	-0.30	}$ & $	6.1	\pm	0.6	$ & $	12.70	_{	-3.77	}^{+	1.81	}$ & $	15.8	^{+	5.0	}_{	-1.4	}$ \\
2004ao	&	Ib	&	8.43	& $	0.58	^{+	0.26	}_{	-0.18	}$ & $	. . .			$ & $	. . .					$ & $	. . .					$ \\
2004dg	&	IIP	&	8.53	& $	0.74	^{+	0.33	}_{	-0.23	}$ & $	181.0	\pm	28.3	$ & $	5.85	_{	-0.07	}^{+	0.07	}$ & $	33.0	^{+	0.7	}_{	-0.7	}$ \\
2004dj*	&	IIP	&	8.17	& $	0.32	^{+	0.14	}_{	-0.10	}$ & $	. . .			$ & $	. . .					$ & $	. . .					$ \\
2004gt*	&	Ic	&	8.53	& $	0.75	^{+	0.33	}_{	-0.23	}$ & $	209.6	\pm	10.5	$ & $	5.78	_{	-0.03	}^{+	0.03	}$ & $	33.7	^{+	0.2	}_{	-0.3	}$ \\
2005at	&	Ic	&	8.61	& $	0.89	^{+	0.40	}_{	-0.27	}$ & $	6.6	\pm	0.9	$ & $	12.45	_{	-0.13	}^{+	0.18	}$ & $	16.1	^{+	0.1	}_{	-0.2	}$ \\
2005ay*	&	IIP	&	8.48	& $	0.65	^{+	0.29	}_{	-0.20	}$ & $	310.8	\pm	52.8	$ & $	5.55	_{	-0.12	}^{+	0.10	}$ & $	35.9	^{+	1.2	}_{	-1.0	}$ \\
2006ca	&	II	&	8.46	& $	0.63	^{+	0.28	}_{	-0.19	}$ & $	181.6	\pm	28.8	$ & $	5.71	_{	-0.19	}^{+	0.23	}$ & $	37.0	^{+	1.4	}_{	-1.7	}$ \\
2006my	&	IIP	&	8.61	& $	0.89	^{+	0.40	}_{	-0.27	}$ & $	6.4	\pm	0.9	$ & $	12.49	_{	-0.16	}^{+	0.19	}$ & $	16.0	^{+	0.2	}_{	-0.2	}$ \\
2007aa	&	IIP	&	8.50	& $	0.68	^{+	0.31	}_{	-0.21	}$ & $	40.5	\pm	6.0	$ & $	7.57	_{	-0.30	}^{+	0.29	}$ & $	27.4	^{+	0.8	}_{	-0.9	}$ \\
2007gr*	&	Ic	&	8.60	& $	0.87	^{+	0.39	}_{	-0.27	}$ & $	15.6	\pm	1.0	$ & $	7.84	_{	-0.25	}^{+	0.75	}$ & $	24.4	^{+	0.8	}_{	-2.5	}$ \\
2007it	&	II	&	8.51	& $	0.71	^{+	0.32	}_{	-0.22	}$ & $	598.9	\pm	69.6	$ & $	3.69	_{	-0.25	}^{+	0.19	}$ & $	84.2	^{+	13.5	}_{	-10.3	}$ \\
2007oc	&	IIP	&	8.47	& $	0.64	^{+	0.29	}_{	-0.20	}$ & $	20.6	\pm	4.5	$ & $	9.50	_{	-0.83	}^{+	1.06	}$ & $	21.7	^{+	2.4	}_{	-2.2	}$ \\
2007Y	&	Ib	&	8.39	& $	0.54	^{+	0.24	}_{	-0.17	}$ & $	49.5	\pm	12.1	$ & $	6.43	_{	0.49	}^{+	1.31	}$ & $	29.1	^{+	0.1	}_{	-2.2	}$ \\
2008bk*	&	IIP	&	. . .	& $	. . .					$ & $	. . .			$ & $	. . .					$ & $	. . .					$ \\
2008bo*	&	IIb	&	8.61	& $	0.90	^{+	0.40	}_{	-0.28	}$ & $	3.1	\pm	0.5	$ & $	13.50	_{	-0.40	}^{+	1.03	}$ & $	14.9	^{+	0.4	}_{	-0.5	}$ \\
2009bb	&	Ic	&	8.49	& $	0.67	^{+	0.30	}_{	-0.21	}$ & $	311.8	\pm	31.7	$ & $	5.55	_{	-0.07	}^{+	0.05	}$ & $	35.9	^{+	0.7	}_{	-0.5	}$ \\
2009dq	&	IIb	&	8.55	& $	0.77	^{+	0.34	}_{	-0.24	}$ & $	82.2	\pm	9.5	$ & $	6.24	_{	-0.05	}^{+	0.05	}$ & $	29.7	^{+	0.2	}_{	-0.2	}$ \\
2009em*	&	Ic	&	8.43	& $	0.58	^{+	0.26	}_{	-0.18	}$ & $	26.9	\pm	2.7	$ & $	8.56	_{	-0.09	}^{+	0.30	}$ & $	24.5	^{+	0.4	}_{	-1.0	}$ \\
2009H	&	II	&	8.51	& $	0.71	^{+	0.31	}_{	-0.22	}$ & $	. . .			$ & $	. . .					$ & $	. . .					$ \\
2009hd*	&	IIL	&	8.53	& $	0.75	^{+	0.33	}_{	-0.23	}$ & $	57.8	\pm	9.8	$ & $	6.37	_{	-0.06	}^{+	0.08	}$ & $	29.3	^{+	0.2	}_{	-0.1	}$ \\
2009ib	&	IIP	&	8.44	& $	0.60	^{+	0.27	}_{	-0.19	}$ & $	193.1	\pm	23.3	$ & $	5.64	_{	-0.16	}^{+	0.18	}$ & $	37.6	^{+	1.1	}_{	-1.4	}$ \\
2009ip	&	IIn?	&	. . .	& $	. . .					$ & $	. . .			$ & $	. . .					$ & $	. . .					$ \\
2009jf*	&	Ib	&	8.46	& $	0.63	^{+	0.28	}_{	-0.19	}$ & $	1.4	\pm	0.1	$ & $	18.20	_{	-0.52	}^{+	0.77	}$ & $	11.7	^{+	0.3	}_{	-0.5	}$ \\
2009kr*	&	IIL	&	8.52	& $	0.72	^{+	0.32	}_{	-0.22	}$ & $	973.8	\pm	83.8	$ & $	3.26	_{	-0.04	}^{+	0.05	}$ & $	117.0	^{+	5.0	}_{	-6.0	}$ \\
2009ls	&	II	&	8.55	& $	0.78	^{+	0.35	}_{	-0.24	}$ & $	. . .			$ & $	. . .					$ & $	. . .					$ \\
2009md	&	IIP	&	8.51	& $	0.71	^{+	0.32	}_{	-0.22	}$ & $	13.3	\pm	1.7	$ & $	10.96	_{	-2.39	}^{+	0.27	}$ & $	17.9	^{+	4.1	}_{	-0.4	}$ \\
2009N	&	IIP	&	8.38	& $	0.53	^{+	0.23	}_{	-0.16	}$ & $	2.5	\pm	1.1	$ & $	19.68	_{	-4.80	}^{+	2.80	}$ & $	12.7	^{+	2.3	}_{	-1.0	}$ \\
2010F	&	IIP	&	8.53	& $	0.73	^{+	0.33	}_{	-0.23	}$ & $	26.2	\pm	2.9	$ & $	6.89	_{	-0.15	}^{+	0.34	}$ & $	27.5	^{+	0.5	}_{	-1.1	}$ \\
2011fh	&	IIn	&	8.35	& $	0.49	^{+	0.22	}_{	-0.15	}$ & $	. . .			$ & $	. . .					$ & $	. . .					$ \\
2011gv	&	IIP	&	8.53	& $	0.74	^{+	0.33	}_{	-0.23	}$ & $	81.0	\pm	12.0	$ & $	6.24	_{	-0.05	}^{+	0.07	}$ & $	29.7	^{+	0.2	}_{	-0.2	}$ \\
2011jm	&	Ic	&	8.12	& $	0.29	^{+	0.13	}_{	-0.09	}$ & $	33.8	\pm	3.6	$ & $	10.23	_{	-0.21	}^{+	0.17	}$ & $	20.7	^{+	0.6	}_{	-0.5	}$ \\
2012A	&	IIP	&	8.13	& $	0.29	^{+	0.13	}_{	-0.09	}$ & $	151.8	\pm	18.1	$ & $	6.52	_{	-0.17	}^{+	0.21	}$ & $	32.4	^{+	1.2	}_{	-1.5	}$ \\
2012au	&	Ib	&	8.51	& $	0.71	^{+	0.31	}_{	-0.22	}$ & $	124.0	\pm	14.5	$ & $	6.02	_{	-0.06	}^{+	0.09	}$ & $	31.3	^{+	0.6	}_{	-1.0	}$ \\
2012aw	&	IIP	&	. . .	& $	. . .					$ & $	. . .			$ & $	. . .					$ & $	. . .					$ \\
2012cw	&	Ic	&	. . .	& $	. . .					$ & $	. . .			$ & $	. . .					$ & $	. . .					$ \\
2012ec	&	IIP	&	8.45	& $	0.62	^{+	0.28	}_{	-0.19	}$ & $	324.9	\pm	36.3	$ & $	5.04	_{	-0.07	}^{+	0.05	}$ & $	45.0	^{+	1.3	}_{	-1.0	}$ \\
2012fh	&	Ic	&	8.48	& $	0.66	^{+	0.30	}_{	-0.20	}$ & $	1633.0	\pm	191.8	$ & $	3.11	_{	-0.44	}^{+	0.05	}$ & $	120.0	^{		}_{		}$ \\
2012ho	&	IIP	&	8.44	& $	0.60	^{+	0.27	}_{	-0.18	}$ & $	42.5	\pm	4.8	$ & $	7.51	_{	-0.29	}^{+	0.22	}$ & $	27.5	^{+	0.9	}_{	-0.6	}$ \\
2012P	&	IIb	&	8.53	& $	0.74	^{+	0.33	}_{	-0.23	}$ & $	181.0	\pm	28.3	$ & $	5.85	_{	-0.07	}^{+	0.07	}$ & $	33.0	^{+	0.7	}_{	-0.7	}$ \\
2013F	&	Ib/c	&	8.58	& $	0.84	^{+	0.37	}_{	-0.26	}$ & $	120.7	\pm	15.0	$ & $	6.00	_{	-0.03	}^{+	0.13	}$ & $	31.5	^{+	0.3	}_{	-1.3	}$ \\
2017ahn	&	II	&	8.52	& $	0.73	^{+	0.33	}_{	-0.22	}$ & $	22.1	\pm	2.9	$ & $	8.17	_{	-1.12	}^{+	1.60	}$ & $	23.3	^{+	3.7	}_{	-4.0	}$ \\
\end{longtable}
\tablefoot{Entries noted with asterisk (*) are from \citet{hk13a,hk13b}, and those without age estimate have metallicity derived from the nearest H~\textsc{ii} region.} \\
\tablefoottext{1}{In \citet{marino13} scale, N2 calibration.  \\
\tablefoottext{2}{Assuming 12+log(O/H)$_\odot = 8.59$ \citep{asplund09} and taking into account 0.16 dex error in the N2 calibration for 12+log(O/H) \citep{marino13}.} \\
}
}

\end{appendix}

\end{document}